\documentclass[%
reprint,
superscriptaddress,
bibnotes,
amsmath,amssymb,
aps,
floatfix,
longbibliography
]{revtex4-1}

\usepackage{qcircuit}
\usepackage[dvipdfmx]{graphicx}
\usepackage{graphicx}
\usepackage{amsmath,amssymb,amsthm,mathrsfs,amsfonts,dsfont}
\usepackage{subfigure, epsfig}
\usepackage{braket}
\usepackage{bm}
\usepackage{enumerate}
\usepackage[colorlinks,linkcolor=blue,citecolor=blue]{hyperref}

\newcommand{\tr}{\mathrm{Tr}}

\newcommand{\black}[1]{\textcolor{black}{#1}}
\newcommand{\mathh}{\mathcal{H}}
\newcommand{\maths}{\mathcal{S}}
\newcommand{\vecc}{\vec{\alpha}}
\newcommand{\Sm}{S^{(M)}}
\newcommand{\Om}{O^{(M)}}
\newcommand{\bbraket}[1]{\braket{\braket{#1}}}

\begin{document}
\title{Generalized quantum subspace expansion}

\author{Nobuyuki Yoshioka}
\email{nyoshioka@ap.t.u-tokyo.ac.jp}
\affiliation{Department of Applied Physics, University of Tokyo, 7-3-1 Hongo, Bunkyo-ku, Tokyo 113-8656, Japan}
\affiliation{Theoretical Quantum Physics Laboratory, RIKEN Cluster for Pioneering Research (CPR), Wako-shi, Saitama 351-0198, Japan}

\author{Hideaki Hakoshima}
\affiliation{Research Center for Emerging Computing Technologies,  National  Institute  of  Advanced  Industrial  Science  and  Technology  (AIST),1-1-1  Umezono,  Tsukuba,  Ibaraki  305-8568,  Japan.}
\affiliation{Center for Quantum Information and Quantum Biology, Osaka University, 1-3 Machikaneyama,Toyonaka, Osaka 560-8531, Japan}

\author{Yuichiro Matsuzaki}
\affiliation{Research Center for Emerging Computing Technologies,  National  Institute  of  Advanced  Industrial  Science  and  Technology  (AIST),1-1-1  Umezono,  Tsukuba,  Ibaraki  305-8568,  Japan.}
\affiliation{NEC-AIST Quantum Technology Cooperative Research Laboratory,
National Institute of Advanced Industrial Science and Technology (AIST), Tsukuba, Ibaraki 305-8568, Japan}

\author{Yuuki Tokunaga}
\affiliation{NTT Computer and Data Science Laboratories, NTT Corporation, Musashino 180-8585, Japan}

\author{Yasunari Suzuki}
\affiliation{NTT Computer and Data Science Laboratories, NTT Corporation, Musashino 180-8585, Japan}
\affiliation{JST, PRESTO, 4-1-8 Honcho, Kawaguchi, Saitama, 332-0012, Japan}

\author{Suguru Endo}
\email{suguru.endou.uc@hco.ntt.co.jp}
\affiliation{NTT Computer and Data Science Laboratories, NTT Corporation, Musashino 180-8585, Japan}

\newcommand{\YM}[1]{\textcolor[rgb]{0.1, 0.5, 0.1}{#1}}
\newcommand{\YMdel}[1]{\textcolor[rgb]{0.1, 0.5, 0.1}{\sout{\textcolor{black}{#1}}}}


\begin{abstract}
One of the major challenges for erroneous quantum computers is undoubtedly the control over the effect of noise.
Considering the rapid growth of available quantum resources that are not fully fault-tolerant, it is crucial to develop practical hardware-friendly quantum error mitigation (QEM) techniques to suppress \black{unwanted} errors.
Here, we propose a novel generalized quantum subspace expansion method which can handle stochastic, coherent, and algorithmic errors in quantum computers.
By fully exploiting the substantially extended subspace, we can efficiently mitigate the noise present in the spectra of a given Hamiltonian, without relying on any information of noise.
The performance of our method is discussed under two highly practical setups: the quantum subspaces are mainly spanned by powers of the noisy state $\rho^m$ and a set of error-boosted states, respectively.
We numerically demonstrate in both situations that we can suppress errors by orders of magnitude, and show that out protocol inherits the advantages of previous error-agnostic QEM techniques as well as overcoming their drawbacks.
\end{abstract}
\maketitle

\emph{Introduction.---} 
Control over computational errors is one of the central problems for the implementation of practical quantum computing algorithms using quantum devices subject to imperfections~\cite{nielsen2002quantum,lidar2013quantum}. 
Towards the goal of achieving fully fault-tolerant computation based on logical operations, the number of required qubits was reduced, and their error rates were improved drastically in the recent years, although the realization of ultimate digital quantum computing is years ahead~\cite{preskill2018quantum}.
Therefore, it is important to ask whether we can establish information processing techniques which exploit the increasing quantum resource {\it without} performing fully-functional error correction.

The quantum error mitigation (QEM) techniques perform post-processing on measurement data (usually expectation values) to eliminate unwanted bias from computation results, in exchange for additional measurement costs~\cite{endo2021hybrid,temme2017error, li2017efficient,endo2018practical,mcardle2019error,bonet2018low,caiMultiexponentialErrorExtrapolation2020,sun2021mitigating,kandala2019error,song2019quantum,zhang2020error,sagastizabal2019experimental}.
One of the most prominent examples is the quasi-probability method~\cite{temme2017error,endo2018practical}. Once the error profile of gate operations is given, stochastic operations are inserted to construct the inverse \textcolor{black}{operations} of each error map so that we can retrieve the computation result for the intended quantum operation. 
However, the characterization of the noise model, e.g., via the gate set tomography, is quite costly and easily deteriorated by noise drift. 

Meanwhile, error-agnostic QEM methods which do not rely on prior knowledge on the error have been proposed: the quantum subspace expansion~(QSE) method~\cite{mcclean_2017,mcclean2020decoding, takeshita_2020, yoshioka2020variational} and the virtual distillation~(VD) method, which is also called exponential error suppression (EES) method~\cite{huggins2020virtual, koczor2020exponential, czarnik2021qubit, huo2021dual}.
In the QSE method, we classically realize a variational subspace spanned by a set of quantum states $\{\ket{\psi_i}\}_i$ as $\ket{\psi}=\sum_i c_i \ket{\psi_i}$, which can be effectively generated via additional measurements and post-processing. 
While the QSE method was initially proposed to compute excited states from a ground state realized on a quantum device, it also contributes to the mitigation of errors.
By construction, the QSE method is well-suited for mitigating coherent errors which may come from insufficient variational optimization, lack of quantum circuit representability, and etc. 
However, it cannot suppress stochastic errors efficiently, since in general we need a linear combination of exponentially many Pauli operators to construct a projector to the error-free subspace~\cite{mcclean_2017,endo2021hybrid}. 
The VD/EES method, on the other hand, is complementary in this sense. 
By applying entangling operations between $M$ identical copies of noisy quantum states $\rho$, we can obtain the error-mitigated expectation value of an observable $O$ as $\braket{O}_{\rm VD}^{(M)}=\tr[O \rho_{\rm VD}^{(M)}]$ with $\rho_{\rm VD}^{(M)}=\rho^M/{\rm Tr}[\rho^M]$, whose fidelity with a dominant eigenvector of $\rho$ exponentially approaches unity. 
Although this method can significantly compensate for stochastic errors, it is entirely vulnerable to coherent errors which distorts the dominant eigenvector. 

\begin{figure*}[t]
\begin{center}
\resizebox{0.9\hsize}{!}{\includegraphics{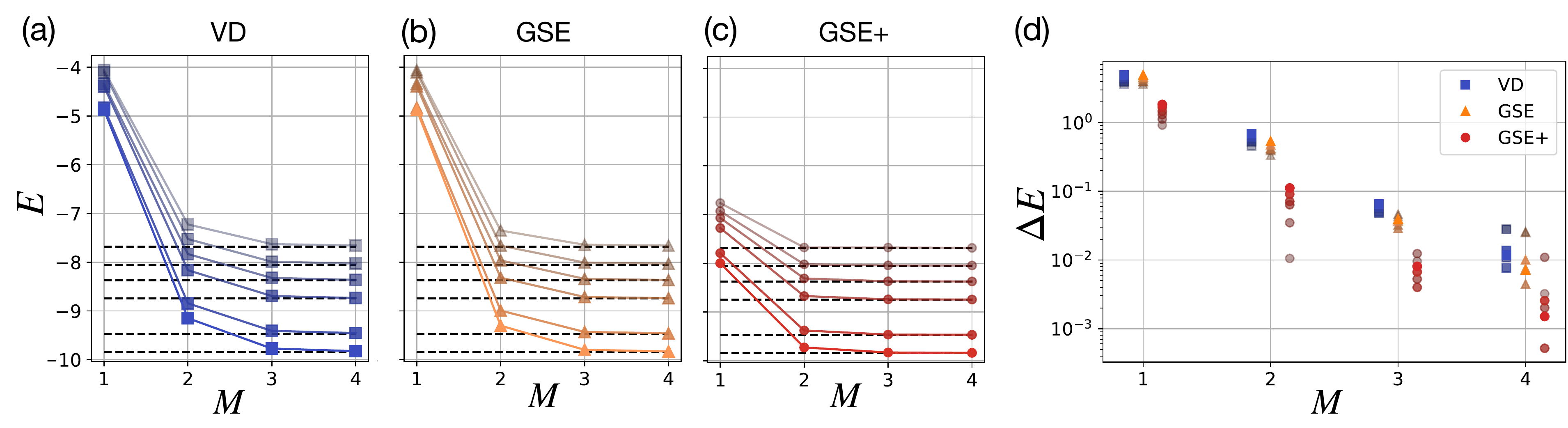}}
\caption{\black{Suppressing errors in 6 lowest eigenstate calculation of one-dimensional transverse-field Ising model by interfering $M$ copies of identical noisy quantum states. 
Eigenenergies computed by (a) VD/EES method, (b) GSE method based on the power subspace, and (c) GSE method with additional bases.
For the power subspace, we take the bases as $\sigma_i = \rho^i~(i=0, 1, ..., \frac{M}{2})$ and $A=I$ for even number of copies $M$, while we take $\sigma_i = \rho^i~(i=0, 1, ..., \frac{M-1}{2})$ and $A=\rho$ for odd $M$'s.
In (c), we additionally include non-Hermite operators $\rho^m H~(m=0, 1,..., \lfloor M/2\rfloor) $. 
(d) The log scale plot of the deviation $\Delta E$ from the exact eigenenergies.
For each eigenstate level $n$, we generate the noisy state $\rho$ by adding depolarizing error after each gate of a variational quantum circuit, whose parameters are optimized by the \black{subspace-search VQE algorithm~\cite{nakanishi2019subspace} to solve an 8-qubit system under $h=1$. }
The depolarizing error rate $p_{\rm dep} $ is taken so that expected number of total error in $\rho$ is given as $N_{\rm tot} = N_{\rm gate} p_{\rm dep}$ where $N_{\rm gate}$ is the number of gates. For all data presented in this figure we set $N_{\rm tot} = 1.5$. }
}
\label{fig:gse}
\end{center}
\end{figure*}

In this work, we propose a unified framework of error-agnostic QEM techniques which we refer to as the generalized quantum subspace expansion~(GSE) method.
The central idea is to extend the notion of quantum subspaces to include general operators that are related to the target noisy quantum state, which allows us to distill the state into an error-mitigated eigenstate of the target Hamiltonian.
We show that the GSE method, which provides a substantial generalization of the QSE method, inherits the advantages of previous error-agnostic QEM techniques as well as overcoming their drawbacks.
This is demonstrated under two practical choices of the subspace.
In the first example, the subspace consisting of powers of a noisy quantum state $\rho^m$ achieves not only the exponential suppression of stochastic errors which is even more efficient than the VD/EES method, but also efficiently mitigates coherent errors.
In the second example, we span the subspace by non-equivalent quantum states corresponding to different noise levels.
Unlike the commonly used error-extrapolation method, the GSE method with the subspace of error-controlled states is quite robust even when the control over noise level is imprecise, and hence highly beneficial to practical applications. 

\emph{Framework of generalized quantum subspace expansion.---} \black{Suppose we obtain a noisy approximation $\rho$ of some desired state, e.g. \black{an eigenstate} of a given Hamiltonian $H$ using the variatioanal quantum eigensolver (VQE) \black{or its variants}~\cite{peruzzo2014variational, kandala2017hardware, nakanishi2019subspace, higgott2019variational, jones2019variational, yoshioka_2020_dvqe,RevModPhys.92.015003,cao2019quantum,cerezo2020variational,bharti2021noisy}}. 
The GSE method uses the following ansatz in the extended subspace to repserent an eigenstate:
\begin{align}
\rho_{\rm EM}=  \frac{P^\dag A P}{\mathrm{Tr}[P^\dag A P]}, \label{Eq:gse_ansatz}
\end{align}
where $P=\sum_i \alpha_i \sigma_i~(\alpha_i \in \mathbb{C})$ is a general operator, $\sigma_i$ is generally a non-Hermite operator, and $A$ is a positive-semidefinite Hermite operator. 
In this paper, we refer to $\sigma_i$ as a base of subspace. 
It is easy to check that $\rho_{\rm EM}$ is a positive-semidefinite Hermite operator whose trace is unity, which ensures that $\rho_{\rm EM}$ corresponds to a physical quantum state. Note that $\sigma_i$ and $A$ can be related to the noisy state $\rho$. 
For example, we can choose $\sigma_i = \rho$ and $A=\rho$; this highlights the crucial difference of the novel GSE method from the conventional QSE~(see Supplementary materials (SM) for more details~\footnote{See Supplementary Materials for more details (URL to be added)}) that it also includes general operators related to quantum states in the expanded subspace. 
To span the most general subspace, we can take \black{a base} as follows,
\begin{equation}
\sigma_i= \sum_k \beta^{(i)}_k \prod_{l=1}^{L_k} U_{l k}^{(i)} \rho_{l k}^{(i)} V_{l k}^{(i)},
\label{Eq: generalsubspaceintro}
\end{equation}
where $\beta^{(i)}_k \in \mathbb{C}$, $\rho^{(i)}_{lk}$ is a quantum state, $U_{l k}^{(i)}$ and $V_{l k}^{(i)}$ are operators that allow for an efficient measurements on quantum computers (e.g. local Pauli operators or unitary operators), and $L_k$ denotes the number of quantum state. See SM for more details~\cite{Note1}.

To obtain the error-mitigated spectra of the Hamiltonian, we 
\black{determine the coefficients $\vec{\alpha} = (\alpha_0, \alpha_1,...)$ by}
solving the following generalized eigenvalue problem~\cite{Note1}:
\begin{equation}
\begin{aligned}
\mathcal{H}\vec{\alpha}&= E \mathcal{S} \vec{\alpha},
\label{Eq:GSE_equation}
\end{aligned}
\end{equation}
where $\mathcal{H}_{ij}=\tr[\sigma_i^\dag A \sigma_j H ]$ and  $\mathcal{S}_{ij}=\tr[\sigma_i^\dag A  \sigma_j]$ with $E$ being the error-mitigated eigenenergy. The coefficients are normalized as $\vec{\alpha}^\dag \maths \vec{\alpha}=1$ to satisfy $\tr[\rho_{\rm EM}]=1$. 
Note that $\mathcal{H}_{ij}$ and $\mathcal{S}_{ij}$ need to be efficiently computed on quantum computers.
Once we find $\vec{\alpha}$ which suffices \black{Eq.~(\ref{Eq:GSE_equation})}, we can compute the error-mitigated expectation value of any observable $O$ as  $\braket{O}=\sum_{ij} \alpha_i^* \alpha_j
\tr[ \sigma_i^\dag A \sigma_j O  ]$.

By implementing the generalized quantum subspaces spanned by Eq.~\eqref{Eq: generalsubspaceintro}, \black{we can efficiently perform error-agnostic QEM}.
To illustrate the significance of our scheme, we will describe slightly more specific but highly practical two subclasses. Due to their features explained thereafter, we refer to the employed subspaces as the {\it power subspace} and {\it fault subspace}, respectively.

\emph{Power subspace.---}
 Let us first restrict the bases of subspace to powers of noisy quantum states as $\sigma_i =\rho^i~(i=0,1,...,m)$ and set $A=I$:
 \begin{equation}
\begin{aligned}
 \rho_{\rm EM} &= \sum_{i,j=0}^m \alpha_i^* \alpha_j \rho^{i+j},
 \label{Eq:unnormalised}
\end{aligned}
\end{equation}
This shows that the error-mitigated state $\rho_{\rm EM}$ is represented as the series expansion of the state $\rho$ as $\rho_{\rm EM}= \sum_{n=0}^{2m} f_n \rho^n$ where $f_n=\sum_{i+j=n} \alpha_i^* \alpha_j$.
Setting $m=1$, for instance, leads to $\rho_{\rm EM}=f_0 I+f_1\rho+f_2 \rho^2$, which clarifies that $\rho_{\rm EM}$ is represented as a polynomial of $\rho$~\footnote{We may alternatively take $A = \rho$ to obtain $\rho_{\rm EM} = f_1 \rho + f_2\rho^2 + f_3 \rho^3$ for $m=1$.}.

\begin{figure}[t]
\begin{center}
\resizebox{0.65\hsize}{!}{\includegraphics{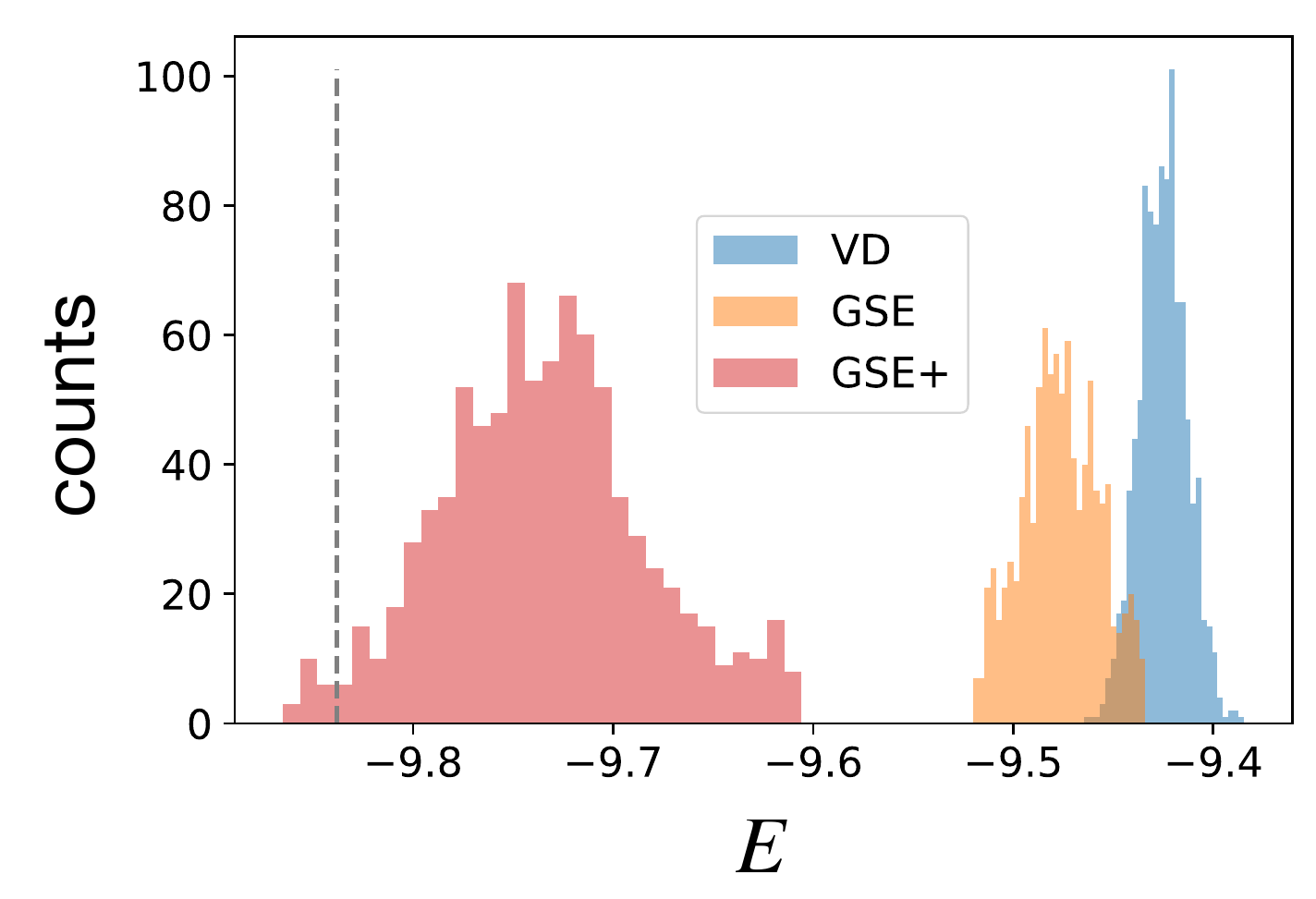}}
\caption{\black{Histograms of ground-state energy estimation by VD/EES (blue), GSE method based on the power subspace (orange), and GSE+ method that includes additional term $\rho H$ included in power-subspace bases \black{(red)} using $M=2$ copies. Here, we take the number of total measurement shots to be $10^{9}$. The gray dotted line indicates the exact ground state energy of 1d TFI model with $N=8$ qubits. }}
\label{fig:power_shotnoise}
\end{center}
\end{figure}

\begin{figure}[t]
\begin{center}
\resizebox{0.65\hsize}{!}{\includegraphics{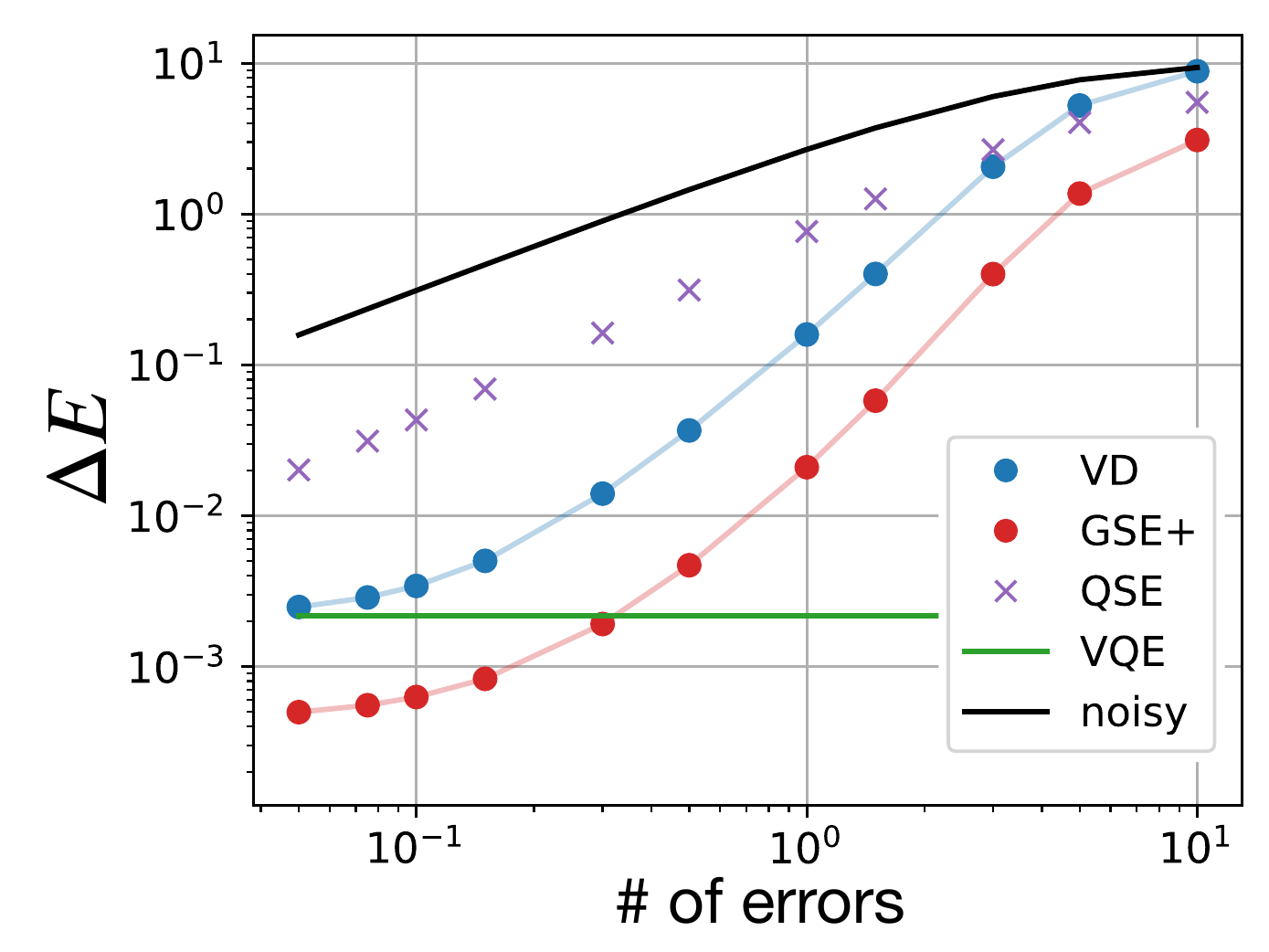}}
\caption{Relationship of the expected number  of errors $N_{\rm tot}$ and the ground-state energy deviation $\Delta E$.
\black{Blue filled circles and red filled circles denote the data from the VD/EES and GSE+ methods using $M=2$ copies of identical noisy quantum states, respectively. \black{Note that GSE+ denotes the GSE method with additional term $\rho H$ included in the bases of subspace $\{\sigma_i\}$.} \black{The purple crosses indicate the ordinary QSE method which corresponds to choosing $A=\rho$ and $\sigma_i \in \{I, H\}$. }
The black and green lines indicate results from the raw noisy quantum state and error-free optimized circuit, respectively.}
While the accuracy by the VD/EES result is bounded by the insufficient expressibility of the variational quantum circuit, the GSE method can reach beyond this limit by further exploring the subspace.
}
\label{fig:gse_pdep}
\end{center}
\end{figure}

It has been pointed out that higher order states themselves are extremely useful~\cite{huggins2020virtual, koczor2020exponential,cai2021quantum}.
By effectively computing the expectation value of an observable corresponding to the state $\rho_{\rm VD}^{(M)}=\rho^M/\tr[\rho^M]~(M=2,3,...)$, we can exponentially suppress the contribution from the non-dominant eigenstates of $\rho$ (See SM for details~\cite{Note1}).
Our key insight is that the non-dominant states will be suppressed even more efficiently by interfering them with each other.
In fact, it is straightforward to see that the power subspace for $A=I$ completely includes $\rho_{\rm VD}^{(2m)}$, and therefore \black{in the case of ground-state simulation} we can always surpass the performance of the VD/EES method when the dominant vector gives good approximation of the ground state~\cite{Note1}.

To illustrate the expected gain by our approach, we numerically demonstrate our algorithm. Figure~\ref{fig:gse} shows the results for 6 lowest eigenstates of the one-dimensional transverse-field Ising (1d TFI) model, whose Hamiltonian is given as $H=-\sum_r Z_r Z_{r+1} + h\sum_r X_r$ where $X_r$ and $Z_r$ denote the $x$-and $z$-components of the Pauli matrix acting on the $r$-th site and $h$ is the amplitude of the transverse magnetic field. We set $h=1$ in the following.
It is clear from Fig.~\ref{fig:gse} that both the VD/EES method and our GSE method yields exponential suppression of error with respect to the number of copies $M$.
Moreover, the interference with non-dominant states in $\rho$ yields quicker convergence of the expectation value $\tr[\rho_{\rm EM} H]$ towards the \black{exact values}; this is further boosted by including additional operators such as $\rho^m H$ to the bases, which is discriminated as GSE+ method in the figures. 
\black{While we observe a trade-off between the accuracy and estimation variance as shown in Fig.~\ref{fig:power_shotnoise}, the greater suppression in GSE/GSE+ method gives us an advantage when the measurement resources are not too scarce.} 
\black{Such a gain in the performance is found not only in the energy, but also measures such as the fidelity and trace distance (See SM for details~\cite{Note1}).}

Now, let us further analyze the effect of the crucial obstacle for the previous exponential error suppression techniques---the coherent errors. 
It has been pointed out in Refs.~\cite{huggins2020virtual, koczor2020exponential,koczor2021dominant} that the stochastic gate errors themselves may cause a deviation of the dominant vector, which is called the coherent mismatch. 
In addition, there are numerous other sources that give rise to the coherent errors, e.g., \black{restrictions on the variational ansatz structure of quantum states due to experimental limitations}.
In this regard, we interestingly find that our method provides a significant improvement over previous methods, since the expressibility of quantum states can be enhanced effectively by the subspace.

Figure~\ref{fig:gse_pdep} shows the result for numerical simulations \black{focused on the ground state} to support our findings. 
While the accuracy of the raw noisy state and the conventional QSE method scales only linearly with respect to $N_{\rm tot}$, both the VD/EES and GSE methods using two copies of $\rho$ provide quadratic suppression in the noisy regime. 
However, the difference of two methods is highlighted in the low-error regime, in which the accuracy of the VD/EES method is bounded by the the performance of the original VQE simulation. 
Namely, when the ideal quantum circuit is not powerful enough and involves algorithmic error, we cannot remedy the shortage by merely restoring the dominant vector. 
In sharp contrast, our method is capable of eliminating such unwanted errors. 

It is important to remark that the required number of measurements for the GSE method scales quadratically with respect to the desired accuracy, just as in the usual quantum measurements (See SM for details~\cite{Note1}).
When the dominant vector of $\rho$ gives a good approximation of the ground state, this is mainly accounted for by the sampling cost rooting from higher powers $\rho^M$.

\emph{Fault subspace.---} 
Now we proceed to another practical subclass of the \black{GSE framework} that employs non-identical quantum states to span the quantum subspace. 
Here, the error-agnostic QEM is realized by utilizing quantum states from different noise levels, and hence refer to the subspace as the {\it fault subspace}; we take $\sigma_i = \rho(\lambda_i \epsilon)$ where $\epsilon$ is the unit of the controlled error (e.g., infidelity per gate) and $\lambda_i \geq 1$ determines the actual error level. For instance, we consider an error-mitigated state as follows:
\begin{equation}
    \rho_{\rm EM} = \sum_{ij}\alpha_i^* \alpha_j \rho(\lambda_i \epsilon) \rho(\lambda_j \epsilon),
    \label{Eq:fault}
\end{equation}
where we have set $A=I$ and $\sigma_i = \rho(\lambda_i\epsilon)$.
We may also extend the fault subspace to include high orders $\rho^m(\lambda_i \epsilon)~(m\geq 2)$ or operators $U_l^{(i)}$ and $V_l^{(i)}.$ 

\begin{figure}[t]
\begin{center}
\resizebox{0.75\hsize}{!}{\includegraphics{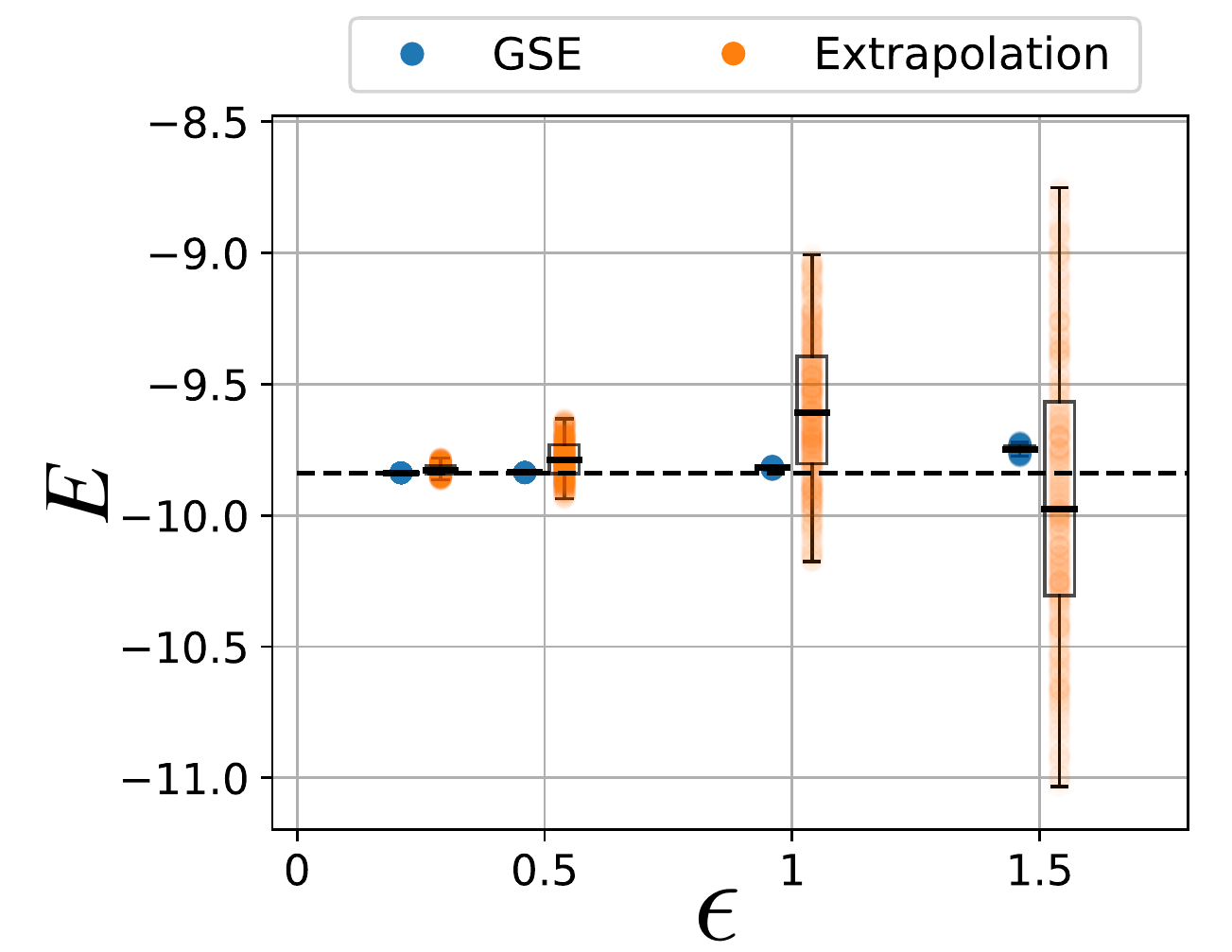}}
\caption{Influence of fluctuation in the stretch factor $\lambda_i$. 
The blue and orange points denote the results from the GSE method using fault subspace and the extrapolation method for the VD/EES calculation using $M=2$ copies, respectively.
It can be clearly observed that the extrapolation method under uncertain noise control yields both systematic deviation and increased variance.
For each error unit $\epsilon$, we generate 500 sets of noisy quantum states $\rho(\hat{\lambda}_i \epsilon)$ where $\hat{\lambda}_i = \lambda_i + \mathcal{N}(0, \lambda_i \epsilon\sigma^2)$ for $\lambda_i \in \{1, 2, 3\}$ and $\sigma = 0.1$.
We and assume that each Pauli term is estimated without any shot-noise.
}
\label{fig:fse}
\end{center}
\end{figure}

The concept of the fault subspace is closely related to the celebrated error-extrapolation method~\cite{li2017efficient,temme2017error}.
\black{In the error-extrapolation method, one estimates the zero-noise limit of the expectation value of a given observable $O$ based on results at $n+1$ noise levels $\braket{O(\lambda_i \epsilon)}=\tr[\rho(\lambda_i \epsilon) O]$.
The estimated computation result is given as
$O^* = \sum_{i=0}^n \beta_i \braket{O(\lambda_i \epsilon)}+\mathcal{O}(\epsilon^{n+1})$
where $\beta_i \in \mathbb{R}$, $\sum_{i=0}^n \beta_i=1$ and $\sum_{i=0}^n \beta_i \lambda_i ^k=0$ for $k=1,2, ..., n$ (See SM for details~\cite{Note1}).
This implies that the error-extrapolation method constructs an effective density matrix as $\rho_{\rm ex}=\sum_{i=0} ^n \beta_i \rho(\lambda_i \epsilon)$.}

Due to its simplicity and practicality, the extrapolation method has been investigated widely both theoretically and experimentally.
However, the extrapolation is based on a highly nontrivial assumption that the noise level can be accurately controlled (e.g. by extending the gate execution duration).
Moreover, since the extrapolation is a purely mathematical operation that does not take any physical constraint into account, it may produce unphysical results even if the measurement is done perfectly, \textcolor{black}{e.g., $\rho_{\rm ex}$ can be a unphysical state whose eigenvalues can be negative.} 

The GSE method using the fault subspace can solve the above problems. First, the results obtained from the GSE method corresponds to a physical density matrix. Second, the GSE method using the fault subspace does not rely on the accurate knowledge of noise levels. This is because the GSE method simply aims to construct a truncated Hilbert space so that the lowest eigenstate is included. It suffices to employ bases that are not identical to each other, while the choice of error levels may affect the practical efficiency.

\black{As a demonstration, we numerically investigate the ground state of 1d TFI model assuming that the control over the noise level is imperfect (See SM for simulation of excited states~\cite{Note1}).}
Here, we consider three noise levels $\rho_i = \rho(\hat{\lambda}_i \epsilon)$ where $\hat{\lambda}_i = \lambda_i + \mathcal{N}(0, \lambda_i \epsilon\sigma^2)$ for $\lambda_i \in \{1, 2, 3\}$ and variance $\sigma^2$. 
The energy at the zero-noise limit is estimated by the Richardson extrapolation for each set of data $\hat{\mathcal{D}} = \{(\lambda_i, \tr[H \rho^2(\hat{\lambda}_i \epsilon)]/\tr[ \rho^2(\hat{\lambda}_i \epsilon)]\}.$ (See SM for details~\cite{Note1}). 
The extrapolated value fluctuates due to the random realization of $\hat{\lambda}_i$, which does not affect the GSE method almost at all.
We highlight this contrast in Fig.~\ref{fig:fse}. 
\black{Due to the stability, the GSE method is suitable for experiments on quantum devices.}

\emph{Summary and Outlook.---}
We have proposed a generalized quantum subspace expansion which unifies the advantages of previously reported error-agnostic methods and furthermore overcomes their drawbacks.
As a practical demonstration, we have first discussed to include powers of the noisy quantum state $\rho^m$ in the base of the subspace. 
This does not only provide the exponential suppression of stochastic error which is even more efficient than the VD/EES method, but it also eliminates the coherent errors of the dominant vector.
In the second strategy, we have presented a method that spans the subspace using quantum states with various noise levels.
Unlike the commonly used error-extrapolation technique, the GSE method exhibits robust performance even when the control over noise level is imprecise. 

There are several future directions. 
First, an efficient combination of the proposed scheme and other QEM methods is worth exploring. For example, we can combine quasi-probability method with the proposed method to suppress the bias of error-mitigated expectation values due to finite characterization errors.  We also expect that exploiting symmetry of the system in the subspace~ \cite{mcclean2020decoding,cai2021quantum} will also improve the computational accuracy. 
Second, our method is not restricted to near-term quantum computing, but may help improve computational accuracy even in the fault-tolerant quantum computing  regimes, when problems of interest involves calculation of eigenspectra.
Namely, we may apply the proposed method to mitigate the effect of errors due to decoding of logical qubits or insufficient number of T-gates without any characterization. This is in contrast with the previous works based on the quasi-probability method~\cite{suzuki2020quantum,piveteau2021error,lostaglio2021error,xiong2020sampling}.
The study of suitable subspace in our GSE framework is also important in future works.

\emph{Acknowledgements.---}
We thank fruitful discussions with Zhenyu Cai, B\'{a}lint Koczor and Kosuke Mitarai.
This work was supported by Leading Initiative for Excellent Young Researchers MEXT Japan and JST presto (Grant No. JPMJPR1919) Japan.
This paper was partly based on results obtained from a project, JPNP16007, commissioned by the New Energy and Industrial Technology Development Organization (NEDO), Japan. This work is supported by PRESTO, JST, Grant No.\,JPMJPR1916, JPMJPR2114, JPMJPR2119; ERATO, JST, Grant No.\,JPMJER1601; CREST, JST, Grant No.\,JPMJCR1771; MEXT Q-LEAP Grant No.\,JPMXS0120319794 and JPMXS0118068682. This work also was supported by JST [Moonshot R\&D][Grant Number JPMJMS2061].
Part of numerical calculations were performed using Qulacs~\cite{suzuki2020qulacs} and QuTiP~\cite{qutip}.

\emph{Note added.---}
Shortly after completion of this work, the reference ~\cite{xiong2021quantum} appeared as a preprint, which consider a method similar to the power subspace corresponding to Eq.~\eqref{Eq:unnormalised}.

\let\oldaddcontentsline\addcontentsline
\renewcommand{\addcontentsline}[3]{}
\bibliography{bib_GSE.bib}
\let\addcontentsline\oldaddcontentsline
\onecolumngrid

\clearpage
\begin{center}
	\Large
	\textbf{Supplementary Materials for: Generalized quantum subspace expansion}
\end{center}

\setcounter{section}{0}
\setcounter{equation}{0}
\setcounter{figure}{0}
\setcounter{table}{0}
\renewcommand{\thesection}{S\arabic{section}}
\renewcommand{\theequation}{S\arabic{equation}}
\renewcommand{\thefigure}{S\arabic{figure}}
\renewcommand{\thetable}{S\arabic{table}}

\addtocontents{toc}{\protect\setcounter{tocdepth}{0}}
{
\hypersetup{linkcolor=blue}
\tableofcontents
}

\section{Quantum subspace expansion method}
\black{In this section, we provide a concise review on the quantum subspace expansion (QSE) method~\cite{mcclean_2017, mcclean2020decoding, yoshioka2020variational}. 
In short, the QSE method can be understood as a post-processing technique that allows one to further explore the Hilbert space in variational simulation.}
\black{Namely, given a set of (non-orthogonal) quantum states $\{\ket{\psi_i}\}_i$, one considers an effective ansatz given by a linear combination}
\begin{eqnarray}
    \ket{\widetilde{\psi}} = \sum_i c_i \ket{\psi_i}, \label{eq:qse_ansatz}
\end{eqnarray}
\black{
whose coefficients $\vec{c} = [...c_i...]^T$ are determined so that a desired property of $\ket{\widetilde{\psi}}$ is optimized.}

\subsection{QSE method with energy-based variational principle}\label{subsec:qse_en}
\black{A common strategy to simulate the ground state of a given Hamiltonian $H$ is to employ an energy-based variational principle, which is referred to as the Ritz variational principle in literature~\cite{Ritz_1908}. We aim to minimize the following cost function:}
\begin{eqnarray}
\mathcal{L} &=& \braket{\widetilde{\psi}|H|\widetilde{\psi}} - \lambda(\braket{\widetilde{\psi}|\widetilde{\psi}} - 1),\\
&=& \sum_{ij} c_i^* c_j \braket{\psi_i | H |\psi_j} - \lambda\left(\sum_{ij} c_i^* c_j \braket{\psi_i| \psi_j} - 1\right)
\end{eqnarray}
\black{where $\lambda$ is the Lagrange multiplier introduced to restrict the norm of the ansatz~\eqref{eq:qse_ansatz} to be unity.
From the stationary condition $\partial_{c_i^*}\mathcal{L}=0$, we straightforwardly obtain the following generalized eigenvalue problem,
\begin{eqnarray}
    \widetilde{\mathcal{H}} \vec{c} = \lambda \widetilde{\mathcal{S}} \vec{c},\label{eqapp:qse}
\end{eqnarray}
where $\widetilde{\mathcal{H}}_{ij} = \braket{\psi_i|H|\psi_j}$ and $\widetilde{\mathcal{S}}_{ij} = \braket{\psi_i|\psi_j}$ with $\lambda$ yielding the minimal energy achievable within the subspace spanned by $\{\ket{\psi}\}_i$.
Here, we normalize the coefficients to satisfy $\vec{c}^\dag \mathcal{S} \vec{c} = 1$, which directly follows from $\partial_{\lambda}\mathcal{L}=0$.
To stabilize the computation, we further cut off eigenvalues of the metric $\mathcal{S}$ that are below some threshold $\epsilon$~\cite{mcclean_2017, epperly2021theory}.
It is worth mentioning that, while the Ritz variational principle itself is designed for the ground state, other eigenstates also satisfy the stationary condition $\partial_{c_i^*} \mathcal{L} = \partial_{\lambda} \mathcal{L} = 0$ and therefore can be obtained from Eq.~\eqref{eqapp:qse}, if the subspace is properly included.
If needed, one may further compute the energy variance as $\vec{\alpha}^\dag \widetilde{V} \vec{\alpha}/\vec{\alpha}^\dag \widetilde{S} \vec{\alpha}$ where $\widetilde{V}_{ij} = \braket{\psi_i|(H-E)^2|\psi_j}$ to obtain the estimation error of the energy~\cite{weinstein1934modified, imoto2021improving}.
}

\black{
We remark that the idea of extending the variational ansatz after stochastic/numerical optimization was already developed in the field of classical simulation. 
For example, one of the most efficient choices of subspace to suppress algorithmic errors in eigenvalue problem is the Krylov subspace $\{H^i \ket{\psi_0}\}_i$, which motivated us to choose the bases of subspace as $\{I, H\}$ for the calculation in Fig.~3.
However, as was initially pointed out by Ref.~\cite{mcclean_2017}, one of the most important feature of the subspace method in the context of quantum simulation is the suppression of hardware errors.
To the best of our knowledge, there is no unified understanding on how to construct bases that efficiently suppress both algorithmic and hardware errors.
}

\subsection{QSE method with variance-based variational principle}
\black{
One may employ an alternative variational principle to compute the eigenspectra of a given Hamiltonian by utilizing the property of an eigenstate that the energy variance $\braket{H^2} - \braket{H}^2$ shall be zero. 
Following the knowledge of classical variational Monte Carlo simulation~\cite{umrigar_1988, cuzzocrea_2020}, here we define the cost function as 
\begin{eqnarray}
\mathcal{L} = \braket{\widetilde{\psi}|(H - \omega)^2|\widetilde{\psi}} - \lambda(\braket{\widetilde{\psi}|\widetilde{\psi}} - 1),\label{eqapp:cost_var}
\end{eqnarray}
where $\omega$ is an initial guess of the target eigenenergy that may be either fixed or updated iteratively until convergence.
In parallel to the case for energy-based variational principle, we obtain the following:
\begin{eqnarray}
    \widetilde{V}\vec{c} = \lambda \widetilde{S}\vec{c},\label{eqapp:qse_var}
\end{eqnarray}
where $\widetilde{V}_{ij} = \braket{\psi_i|(H-\omega)^2|\psi_j}$. 
After solving Eq.~\eqref{eqapp:qse_var} and choosing the optimal $\vec{c}$ that yields the smallest variance, we can compute the eigenenergy as $E = \vec{c}^\dag \widetilde{H}\vec{c}/\vec{c}^\dag \widetilde{S}\vec{c}$.
As we have mentioned in \S\ref{subsec:qse_en}, the energy variance provides an upper-bound of an estimation error of the energy.
}

\black{As we further mention in Sec.~\ref{sec:gse_exc}, in the current paper we initially set $\omega$ to the energy obtained from the VD/EES method, which is necessarily computed in GSE methods based on the power or fault subspaces. After solving Eq.~\eqref{eqapp:qse_var} and computing the energy, we replace $\omega$ with $E$ and solve the generalized eigenvalue problem again. Note that this step does not require additional measurement on quantum computers; the cost is entirely that of classical computation. }

\section{Virtual distillation or exponential error suppression method}\label{sec:virtualdistillation}  
\begin{figure}[h!]
\begin{align*}
\Qcircuit @C=1.2em @R=1.2em {
\lstick{\ket{0}}&\gate{H}&\qw&\qw&\ctrl{1}&\ctrl{2}&\ctrl{1} &\measureD{X}\\
\lstick{\rho}&\qw&\qw&\qw&\qswap&\qw&\gate{P_a}&\qw \\
\lstick{\rho}&\qw&\qw&\qw&\qswap \qwx &\qswap&\qw&\qw\\
\lstick{\rho}&\qw&\qw&\qw&\qw&\qswap \qwx&\qw&\qw \\}
\end{align*}
\caption{A derangement quantum circuit for evaluating $\tr[\rho^M H]=\sum_a f_{a} \tr [\rho^M P_a]$, where we set $M=3$ and the Hamiltonian is decomposed into a linear combination of products of Pauli operators $P_a$ as $H=\sum_{a} f_{a} P_a$.}
\label{circuit:derangement}
\end{figure}
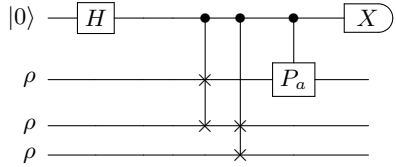

In this section, we review the virtual distillation (VD) method or exponential error suppression (EES) method proposed in Refs.~\cite{huggins2020virtual, koczor2020exponential}.
We consider a noisy state $\rho$, which can be written in terms of the spectral decomposition:
\begin{align}
    \rho=p_0 \ket{\psi_0}\bra{\psi_0}+\sum_{k=1} p_k \ket{\psi_k}\bra{\psi_k},
\end{align}
where we define $\braket{\psi_i| \psi_j}=\delta_{i,j}$ and $p_0 > p_1\ge p_2\ge \cdots \ge0$, and we refer $\ket{\psi_0}$ as a dominant vector.
In that method, we can effectively compute the expectation value of an observable from that dominant vector $\ket{\psi_0}$ with exponentially small error
\begin{align}
    E^{(M)}_{\rm VD}=\frac{\tr[\rho^M H]}{\tr[\rho^M]}&=\frac{{p_0}^M\bra{\psi_0}H\ket{\psi_0}+\sum_{k=1}  (p_k)^M\bra{\psi_k}H\ket{\psi_k}}{{p_0}^M+\sum_{k=1}(p_k)^M}\\
    &=\bra{\psi_0}H\ket{\psi_0}\left[1+\frac{\sum_{k=1} (p_k/p_0)^M(\bra{\psi_k}H\ket{\psi_k}/\bra{\psi_0}H\ket{\psi_0}-1)}{1+\sum_{k=1} (p_k/p_0)^M}\right]\\
    &=E_{\rm dom}\left[1+O\left((p_1/p_0)^{M}\right)\right] \quad (E_{\rm dom}:=\bra{\psi_0}H\ket{\psi_0}),
\end{align}
by measuring the numerator $\tr[\rho^M H]$ and the denominator $\tr[\rho^M]$, respectively. These quantities can be measured by unitary diagonalization~\cite{huggins2020virtual} or indirect (or non-destructive) measurement~\cite{huggins2020virtual, koczor2020exponential,czarnik2021qubit}.
Fig.~\ref{circuit:derangement} shows a controlled derangement quantum circuit for the case of $M=3$ to calculate the numerator $\tr[\rho^M H]$.
We emphasize that when increasing the number of copies $M$, the virtual state $\rho_{\rm VD}^{(M)}=\rho^M/\tr[\rho^M]$ exponentially gets closer to the dominant vector $\ket{\psi_0}$.

The VD/EES methods have the best performance when there are
only stochastic (or orthogonal) errors
that change an ideal state $\ket{\psi_{\rm id}}$ into its orthogonal states.
In this case, we have
$\ket{\psi_0}=\ket{\psi_{\rm id}}$ 
with sufficiently small error rates. 
However, this is not always the case. 
Some type of error causes a change in a state from $\ket{\psi_{\rm id}}$ to non-orthogonal states, and this leads to $\ket{\psi_0}\neq\ket{\psi_{\rm id}}$. The infidelity of these states is called the coherent mismatch $1-|\braket{\psi_0|\psi_{\rm id}}|^2$~\cite{koczor2021dominant}, and the VD/EES method cannot eliminate such
coherent errors in general, although the effect of the coherent errors was investigated through numerical results~\cite{huggins2020virtual, koczor2020exponential} and analytic results~\cite{koczor2021dominant}.

We also remark that even if the state is unphysical, i.e., the state has negative eigenvalues, as long as $|p_0|>|p_k| ~(k=1,2,...)$, the contribution of $|p_k|/|p_0|$ exponentially vanishes as $M$ increases, which implies virtual distillation still works. This ensures that VD/EES method can be applied to error-mitigated unphysical states to further improve the computation accuracy.

\section{Implementation of general subspace}\label{sec:mostgeneral}
\begin{figure}[h!]
\begin{align*}
\Qcircuit @C=0.8em @R=1.2em {
\lstick{\ket{0}}&\gate{H}&\ctrl{1}&\ctrl{2}&\ctrl{1}& \ctrl{2}&\ctrl{3} &\measureD{X, Y}\\
\lstick{\rho}&\qw&\qswap&\qw&\gate{U_1}&\qw&\qw&\qw \\
\lstick{\rho_1}&\qw&\qswap \qwx &\qswap&\qw&\gate{V_1 P_a V_1}&\qw &\qw\\
\lstick{\rho_1}&\qw&\qw&\qswap \qwx&\qw&\qw&\gate{U_1}&\qw \\}
\end{align*}
(a)
\begin{align*}
\Qcircuit @C=0.8em @R=1.2em {
\lstick{\ket{0}}&\gate{H}&\ctrl{2}&\ctrlo{2}&\ctrl{3}&\ctrlo{3}&\ctrl{1}&\ctrl{2}&\ctrl{2} &\measureD{X, Y}\\
\lstick{\rho}&\qw&\qw&\qw&\qw&\qw&\qswap&\qw&\qw&\qw \\
\lstick{\rho_1}&\qw&\gate{U_1}&\gate{V_1^\dag}&\qw&\qw&\qswap \qwx &\qswap&\gate{P_a} &\qw\\
\lstick{\rho_1}&\qw&\qw&\qw&\gate{V_1^\dag}&\gate{U_1}&\qw&\qswap \qwx&\qw&\qw \\}
\end{align*}
(b)
\caption{The quantum circuits for computing $\gamma^{(ij)}_{kk'}$ when $U_1$ 
and
$V_1$ are (a) local Pauli operators or
(b) unitary operators. Since $\gamma^{(ij)}_{kk'}$ is a complex number, we need to calculate the real and imaginary parts by measuring Pauli-X and -Y operators of the ancilla qubit, respectively. $\zeta^{(ij)}_{kk'}$ can be similarly evaluated by excluding the operator $P_a$ from the circuit. }
\label{circuit:mostgeneralsubspace}
\end{figure}
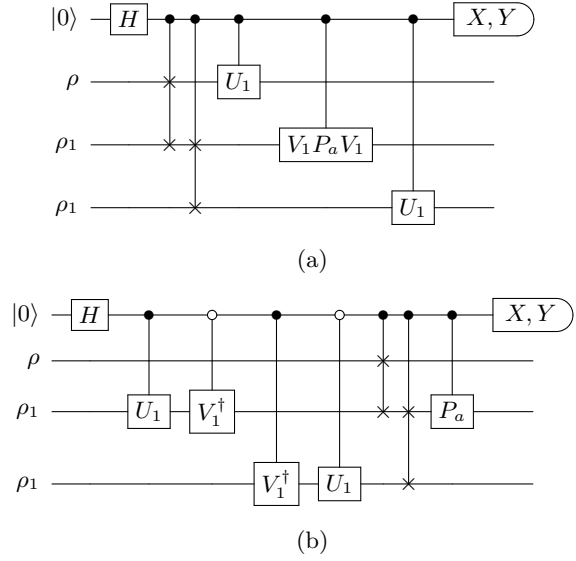

The most general subspace can be implemented using the bases given as follows,
\begin{equation}
\sigma_i= \sum_k \beta^{(i)}_k \prod_{l=1}^{L_k} U_{l k}^{(i)} \rho_{l k}^{(i)} V_{l k}^{(i)},
\label{Eq: generalsubspace}
\end{equation}
where $\beta^{(i)}_k \in \mathbb{C}$, $U_{l k}^{(i)}$ and $V_{l k}^{(i)}$ are general operators, 
$L_k$ denotes the number of quantum state,
 and $\rho^{(i)}_{lk}$ is a quantum state.
Suppose that the Hamiltonian $H$ can
 be decomposed into a linear combination of products of Pauli operators as $H=\sum_a f_a P_a$. The elements $\mathcal{H}_{ij} := \tr[\sigma_i^\dag A \sigma_j H]$ and $\mathcal{S}_{ij} := \tr[\sigma_i^\dag A \sigma_j]$ (which we call  a Hamiltonian in the expanded subspace and an overlap matrix, respectively)
 in the Eq. (\ref{Eq:GSE_equation})
 can be described as:
\begin{equation}
\begin{aligned}
\mathcal{H}_{ij}&= \sum_{k k'} (\beta_k^{(i)})^* \beta_{k'}^{(j)} \gamma^{(ij)}_{kk'}, \nonumber \\
\gamma^{(ij)}_{kk'}&= \sum_a f_a \tr \bigg[\left(\prod_{l=1}^{L_k} (V_{lk}^{(i)})^\dag \rho_{lk}^{(i)} (U_{lk}^{(i)})^\dag\right) A \left(\prod_{l'=1}^{L_{k'}} U_{l'k'}^{(j)} \rho_{l'k'}^{(j)} V_{l'k'}^{(j)}\right) P_a \bigg], \\
\mathcal{S}_{ij}&= \sum_{k k' } (\beta_k^{(i)})^* \beta_{k'}^{(j)}~\zeta^{(ij)}_{kk'},\nonumber \\
\zeta^{(ij)}_{kk'}&= \tr \bigg[\left(\prod_{l=1}^{L_k} (V_{lk}^{(i)})^\dag \rho_{lk}^{(i)} (U_{lk}^{(i)})^\dag \right)A \left(\prod_{l'=1}^{L_{k'}} U_{l'k'}^{(j)} \rho_{l'k'}^{(j)} V_{l'k'}^{(j)}\right) \bigg],
\end{aligned}
\end{equation}
and these
can be calculated on quantum computers by using a modified controlled derangement operator. 
Let us show an example when $A= \rho$ and $L_1=1$, assuming a single subspace given as $\sigma_i = U_1 \rho_1 V_1$.
If both $U_1$ and $V_1$ are local Pauli operators, then the calculation of $\mathcal{H}_{ij}$ can be performed by the quantum circuit shown in Fig.~\ref{circuit:mostgeneralsubspace}(a).
Alternatively, if $U_1$ and $V_1$ are unitary operators, then we can obtain the expectation value from the circuit shown in Fig.~\ref{circuit:mostgeneralsubspace}(b).

\section{Power subspace includes the virtual distillation among the subspace}
We compare the performance of the VD/EES method, $E^{(M)}_{\rm VD}=\frac{\tr[\rho^M H]}{\tr[\rho^M]}$,
 with that of the GSE method using power subspace, $E^{(M)}_{\rm GSE}=\mathrm{min}_{\vec{\alpha}}\tr[\rho_{\rm EM} H]$.
 Since $E^{(M)}_{\rm VD}$ is not a monotonically decreasing function of $M$ in general, we compare $\min_{1\le k\le M} E^{(k)}_{\rm VD}$ with $E^{(M)}_{\rm GSE}$. 
 Actually, we can show $\min_{1\le k\le M} E^{(k)}_{\rm VD}\ge E^{(M)}_{\rm GSE}$.
For this purpose, we can choose $\vec{\alpha}=\vec{\alpha}_k$ ($1\le k \le m$) such that 
\begin{align}
\mathrm{Tr}[\rho_{\rm EM}H]\left.\right|_{\vec{\alpha}=\vec{\alpha}_k}=\begin{cases}
E^{(2k)}_{\rm VD}&(A=I)\\
E^{(2k+1)}_{\rm VD}&(A=\rho)\\
\end{cases}
\label{suppeq:PowerincludeVD}
\end{align}
where we define a vector $\vec{\alpha}_k$:
\begin{align}
\vec{\alpha}_k=\begin{cases}
(\underbrace{0,0,\cdots,0}_{k}, \left(\mathrm{Tr}[\rho^{2k}]\right)^{-1/2},\underbrace{0,\cdots,0}_{m-k})&(A=I)\\
(\underbrace{0,0,\cdots,0}_{k}, \left(\mathrm{Tr}[\rho^{2k+1}]\right)^{-1/2},\underbrace{0,\cdots,0}_{m-k})&(A=\rho).\\
\end{cases}
\end{align}
Note that the $(k+1)$-th component of $\vec{\alpha}_k$ is a non-zero value and all the others are zero.
Eq.~(\ref{suppeq:PowerincludeVD}) means that the power subspace expansion includes the expectation values obtained by the VD/EES methods.
From the above result, we obtain 
\begin{align}
\min_{1\le k\le M} E^{(k)}_{\rm VD}= \min_{1\le k\le M} \mathrm{Tr}[\rho_{\rm EM}H]\left.\right|_{\vec{\alpha}=\vec{\alpha}_k}\ge \mathrm{min}_{\vec{\alpha}}\tr[\rho_{\rm EM} H]= E^{(M)}_{\rm GSE}.
\end{align}
This implies that the convergence of the power subspace expansion is equal or greater than that of the VD/EES methods.

\section{Generalized quantum subspace expansion method for excited states}\label{sec:gse_exc}
\black{
In this section, we further discuss the excited-state simulation by the GSE method.
For the consistency with the main text, here again we consider one-dimensional transverse-field Ising (1d TFI) model under open boundary condition as $H = \sum_r (Z_r Z_{r+1} + h X_r)$ with $h=1$.
In Sec.~\ref{subsec:power_exc} and Sec.~\ref{subsec:fault_exc}, we show that the GSE method with the power subspace and fault subspace can mitigate the errors even for excited states, respectively.}

\black{
Before proceeding to the results by two choice of subspaces, let us summarize the  concrete overall procedure. We focus on simulation of 1d TFI model using the hardware-efficient ansatz whose structure is provided in Sec.~\ref{sec:hardware_efficient}.
\begin{enumerate}
\item[1.] First, we simulate the $K$ lowest eigenstates of the Hamiltonian. Here we use the subspace-search variational quantum eigensolver algorithm (SSVQE)~\cite{nakanishi2019subspace}. The SSVQE algorithm optimizes variational parameters $\theta$ so that the weighted sum of the energy expectation value from multiple orthogonal states is minimized:
\begin{eqnarray}
\theta^* = \mathop{\rm arg~min}\limits_{\theta} \sum_{k=1}^K w_k \braket{\psi_0^{(k)}| U^\dag(\theta) H U(\theta) | \psi_0^{(k)}},
\end{eqnarray}
where initial states satisfy $\braket{\psi_0^{(k)}| \psi_0^{(k')}} = \delta_{kk'}$, weights are positive $w_k>0$, and $K$ is the desired number of excited states. Throughout our paper, we choose the weights to be $w_k = \frac{K-k}{K}$. Note that in our numerical demonstration this optimization is performed assuming that the number of measurement shots is infinite.
\item[(2.] For our numerical demonstration, we introduce single-qubit depolarizing noise after every quantum gate and obtain the noisy state $\rho^{(n)}$ for every
$U(\theta) | \psi_0^{(n)} \rangle$.
This step is skipped when one runs the algorithm on real quantum device. )
\item[3.] Next, for every eigenstate level $n$, we define an effective ansatz as
\begin{eqnarray}
    \rho_{\rm EM}^{(n)} = \frac{P_n^\dag A_n P_n}{\tr [P_n^\dag A_n P_n]} \label{eqapp:gse_exc}, 
\end{eqnarray}
where $P_n = \sum_i \alpha_{ni} \sigma_{ni}$. Then, we solve either of the generalized eigenvalue problems
\begin{eqnarray}
    \mathcal{H}_n\vec{\alpha}_n &=& E^{(n)}_{\rm GSE} \mathcal{S}_n\vec{\alpha}_n,\label{eq:gse_exc_energy}\\
    \mathcal{V}_n\vec{\alpha}_n &=& \lambda_n \mathcal{S}_n\vec{\alpha}_n,\label{eq:gse_exc_variance}
\end{eqnarray}
where $[\mathcal{H}_n]_{ij} = \tr[\sigma_{ni}^\dag A_n \sigma_{nj} H]$, $[\mathcal{V}_n]_{ij} = \tr[\sigma_{ni}^\dag A_n \sigma_{nj} (H-\omega)^2]$, and $[\mathcal{S}_n]_{ij} = \tr[\sigma_{ni}^\dag A_n \sigma_{nj}]$.
For the latter, we initially set the reference energy as $\omega = \tr[(\rho^{(n)})^M H]/\tr[(\rho^{(n)})^M]$, namely the energy obtained from the VD/EES method.
Note that this quantity is necessarily calculated, if one wishes to apply the GSE method.
\item[4.]
\black{Finally, among the multiple solutions of the generalized eigenvalue equation, we choose the one with the smallest variance, which is computed as $\vec{\alpha}_n^\dag \mathcal{V}_n \vec{\alpha}_n/\vec{\alpha}_n^\dag \mathcal{S}_n \vec{\alpha}_n$. One may alternatively compare $E_{\rm GSE}^{(n)} = \vec{\alpha}_n^\dag \mathcal{H}_n \vec{\alpha}_n/\vec{\alpha}_n^\dag \mathcal{S}_n \vec{\alpha}_n$ with some initial guess, e.g., the energy from VD/EES method, and choose the closest value.
It is practically important to remark that, this step is only optional, when one is simulating the lowest (highest) eigenstate using the energy-based variational principle. In such a case, one can typically choose the smallest (largest) solution of Eq.~\eqref{eq:gse_exc_energy}. 
}
\end{enumerate}
}
\black{
Two remarks are in order. 
First, the reference energy $\omega$ introduced in step 3 is a hyperparameter. 
To avoid potential error that may arise from choosing an inappropriate $\omega$, one may iterate steps 3 and 4.
That is, once one solves the generalized eigenvalue problem~\eqref{eq:gse_exc_variance}, one may replace the reference energy with the obtained energy as $\omega= \vec{\alpha}_n^\dag \mathcal{H}_n \vec{\alpha}_n/\vec{\alpha}_n^\dag \mathcal{S}_n \vec{\alpha}_n$ and define a new generalized eigenvalue problem. 
This  procedure can be repeated until the energy/variance converges. 
We remark that this iteration does not require additional measurement on quantum computers; the cost is entirely that of classical computation to solve Eq.~\eqref{eq:gse_exc_energy} or \eqref{eq:gse_exc_variance}.
In the current work, we find the deviation of variance estimation to be sufficiently small, and therefore fix the number of repetition to 2.
Second, to stabilize the computation, we regularize the metric $\mathcal{S}_n$ by cutting off its eigenvalues that are below the threshold $\epsilon = 10^{-8} $~\cite{mcclean_2017, epperly2021theory}.
Note that this value can be adjusted upon request; one may simply solve the generalized eigenvalue problem under various values and check its stability. 
Since this procedure involves only classical computation, it does not require any additional quantum measurement.
}

\begin{figure}[h]
\begin{center}
\resizebox{0.85\columnwidth}{!}{\includegraphics{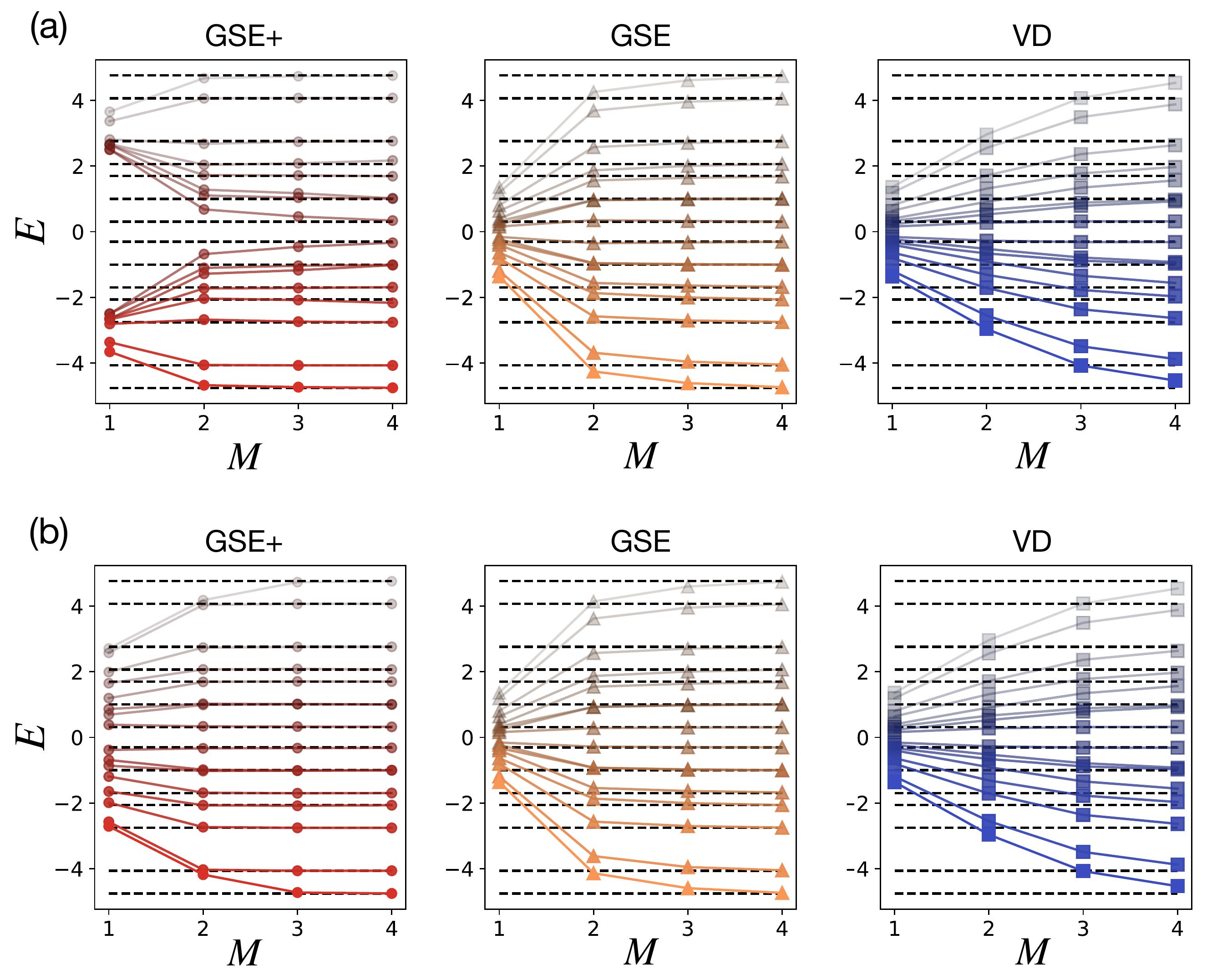}}
\caption{\black{Simulation of all 16 eigenstates of 1d transverse-field Ising model with $N=4$ qubits. (a) Eigenenergies computed from the solution of the generalized eigenvalue problem~\eqref{eq:gse_exc_energy}, which is based on the energy-based variational principle. 
Here, GSE+ indicates that we additionally include non-Hermite term $\rho^m H~(m=0, 1,..., \lfloor M/2\rfloor) $ to the bases of subspace. 
The squares, triangles, and circles denote the data from VD/EES, GSE, and GSE+ methods, respectively.
In (b) we display the result for the computation using Eq.~\eqref{eq:gse_exc_variance}, which follows from the variance-based variational principle. We can see that the GSE method can efficiently mitigate the error over the entire eigenspectra, which is even more significant in the GSE+ method. 
The variational ansatz consists of depth $d=20$ and the total error number is $N_{\rm tot} = 3.0$.}
}
\label{fig:power_subspace_excited_N4}
\end{center}
\end{figure}

\subsection{Power subspace for excited states}\label{subsec:power_exc}
\black{
Here we discuss the performance of the GSE method with the power subspace.
While in the main text we have presented the numerical demonstration for $K=6$ eigenstates in $N=8$ qubit system, here we further attempt to simulate the entire eigenspectra using the GSE method.
The result for simulation of all 16 eigenstates for $N=4$ qubit system is shown in Fig.~\ref{fig:power_subspace_excited_N4}.
\black{We can see from Fig.~\ref{fig:power_subspace_excited_N4}(a) that the GSE methods exhibit greater suppression of errors than the VD/EES method even for excited states.
Another strategy, which we find to be effective for the 1d TFI model as shown in Fig. \ref{fig:power_subspace_excited_N4}(b),
is to employ the variance-based variational principle, or Eq.~\eqref{eq:gse_exc_variance}.
While the performance of the GSE method is similar to the case with the energy-based variational principle, we find improvement in the results of the GSE+ method (especially for $M=3$ and 4).}
\black{We remark that} one may perform more stable calculation by utilizing the symmetry of the system. Namely, one can check whether the solution of Eq.~\eqref{eq:gse_exc_energy} is in desired symmetry sector, e.g., by combining with the symmetry verification methods,
which we leave as a future work.
}

\subsection{Fault subspace for excited states}\label{subsec:fault_exc}
\begin{figure}[t]
\begin{center}
\resizebox{0.85\columnwidth}{!}{\includegraphics{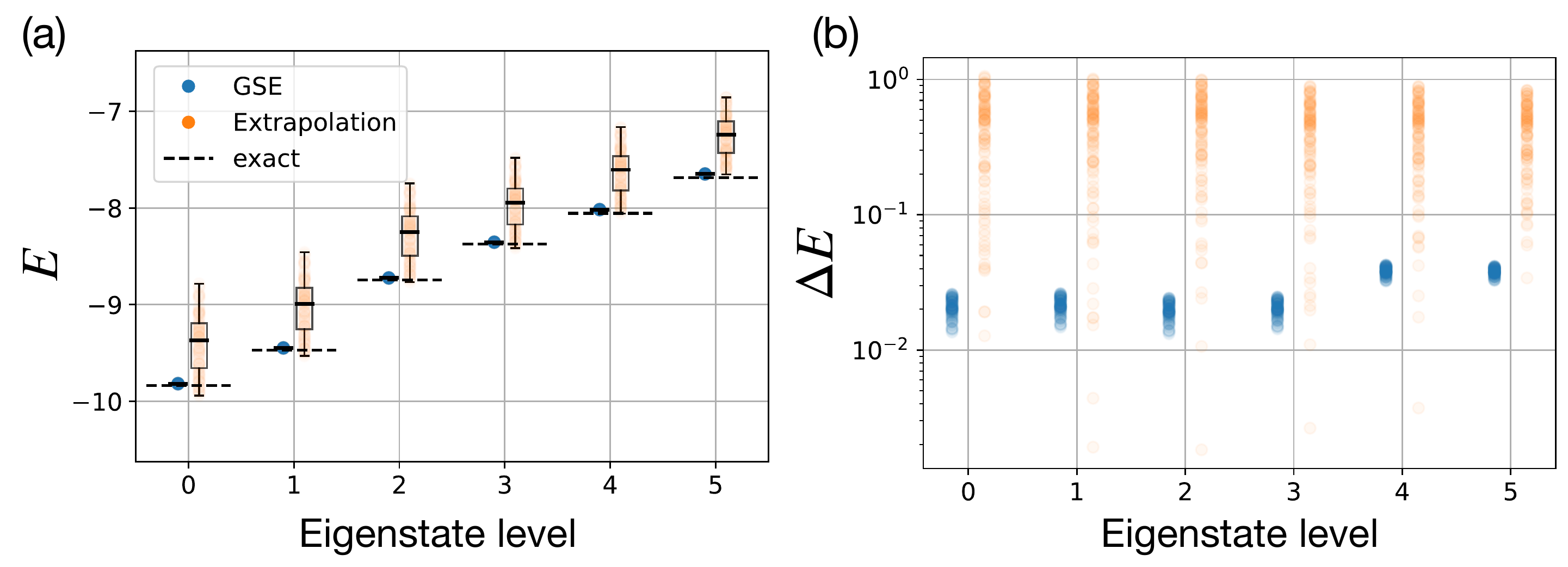}}
\caption{\black{Simulation of 6 minimal eigenstates using the GSE method with the fault subspace (blue) and the error-extrapolation method for the VD/EES calculation using $M=2$ copies (orange) for $N=8$ qubit system. 
Here,  we adopt the energy-based variational principle.
As in the calculation merely for the ground state, we clearly observe that the error-extrapolation method suffers from uncertainty in the stretch factor of errors. Here, we fix the error unit to be $\epsilon=1.0$, and generate 500 sets of noisy quantum states $\rho(\hat{\lambda}_i \epsilon)$ where $\hat{\lambda}_i = \lambda_i + \mathcal{N}(0, \lambda_i \epsilon\sigma^2)$ for $\lambda_i \in \{1, 2, 3\}$ and $\sigma = 0.1$. We assume that each Pauli term is estimated without any shot-noise.
The variational ansatz consists of depth $d=18$.}
}
\label{fig:fault_subspace_excited_N8}
\end{center}
\end{figure}

\black{
Next, we proceed to the simulation of excited states using the fault subspace.
As in the main text, we assume that the stretch factor of the error fluctuates due to noise-control imperfection, which leads the well-known error-extrapolation method~\cite{temme2017error,endo2018practical} to yield unphysical estimation of physical observables (also see Sec.~\ref{sec:error_in_observable}).
Figure.~\ref{fig:fault_subspace_excited_N8}(a) compares the eigenenergy estimation by GSE method using the fault subspace and the error-extrapolation method applied to mitigate errors in the VD method using $M=2$ copies.
Similar to the ground-state simulation as provided in main text (e.g. Fig.~4), the error-extrapolation method suffers from the additional uncertainty where the GSE method with the fault subspace is barely affected.
We can further see from Fig.~\ref{fig:fault_subspace_excited_N8}(b) that there is the unwanted bias introduced in the error-extrapolation method, which results in separation of accuracy in one or two orders.
}

\black{
In Fig.~\ref{fig:fault_subspace_excited_N8} we show the simulation results over the entire eigenspectra for $N=4$ qubit system.
As was done in the simulation using the power subspace, among the solutions obtained by energy-based variational principle (Eq.~\eqref{eq:gse_exc_energy}), we choose the one with the smallest variance.
For the current simulation, we did not observe a large difference in the performance of calculations based on two variational principles.
}

\begin{figure}[t]
\begin{center}
\resizebox{0.75\columnwidth}{!}{\includegraphics{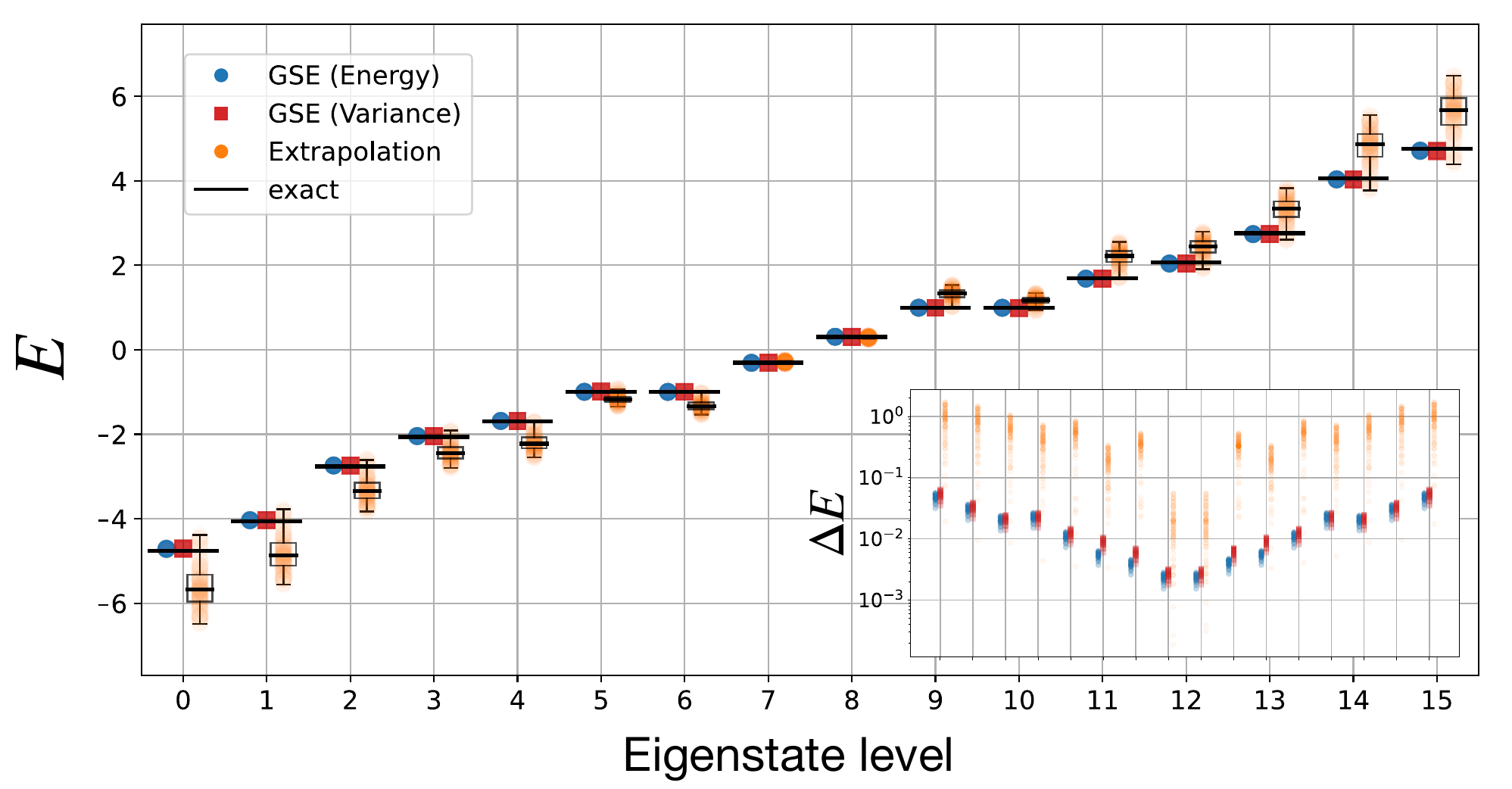}}
\caption{\black{Simulation of all 16 eigenstates of $N=4$ qubit system using the GSE method with the fault subspace with energy-based (blue) and variance-based variational principle (red) and the error-extrapolation method for the VD/EES calculation using $M=2$ copies (orange). The variational ansatz consists of depth $d=20$. \black{The inset shows the log-scale plot of energy error $\Delta E$ for individual eigenstate.}}
}
\label{fig:fault_subspace_excited_N4}
\end{center}
\end{figure}

\section{Additional numerical results on error mitigation performance}\label{sec:error_in_observable}
\black{
In this section, we analyze the performance of the GSE method by comparing the error in physical observables such as two-point correlation and computing  distance measure (such as fidelity or trace distance) from the exact state. In particular, here we focus on the ground-state simulation of 1d transverse-field Ising model.
Note that a physical observable $O$ can be computed as
\begin{eqnarray}
    \braket{O} = \frac{\vec{\alpha}^\dag \widetilde{O}\vec{\alpha}}{\vec{\alpha}^\dag \widetilde{S}\vec{\alpha}},
\end{eqnarray}
where $\widetilde{O}_{ij} = \tr[\sigma_i^\dag A \sigma_j O]$ while $\vec{\alpha}$ is computed by solving the generalized eigenvalue problem.
}

\subsection{Errors of physical observables under power subspace}
\begin{figure}[t]
\begin{center}
\resizebox{0.99\columnwidth}{!}{\includegraphics{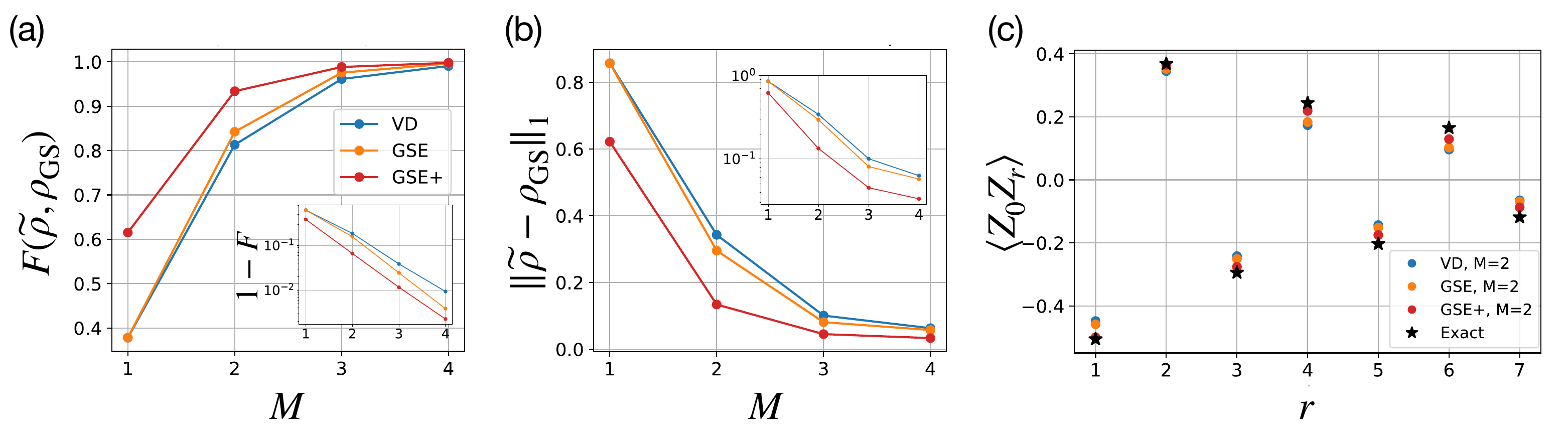}}
\caption{\black{Analysis on the performance of error-mitigated states that are effectively realized by the VD/EES, GSE, and GSE+ methods.  (a) Quantum state fidelity $F$ between the error-mitigated states $\widetilde{\rho}$ and the exact ground state $\rho_{\rm GS}$. The inset shows log plot of the infidelity $1-F$.  (b) Trace distance $\|\widetilde{\rho} - \rho_{\rm GS}\|_1$ with its log plot provided as inset, and (c) expectation values of 2-body correlators $\braket{Z_0 Z_r}$ where $r$ is the site index. In all panels, the blue, orange, and red points denote data by VD/EES, GSE, and GSE+ methods. In (c), we fix the number of copies to $M=2$. The black stars indicate the exact values.
The variational ansatz with depth $d=6$ is optimized via the VQE algorithm. The depolarizing error rate $p_{\rm dep}$ is taken so that expected number of total error in $\rho^{(n)}$ is given as $N_{\rm tot} = N_{\rm gate} p_{\rm dep}$ where $N_{\rm gate}$ is the number of gates. Here, we set $N_{\rm tot} = 2.5$.}
}
\label{fig:obs_error_power}
\end{center}
\end{figure}

\black{
In the main text, we have demonstrated that the GSE method shows stronger suppression of errors in eigenstate energy over the QSE or VD/EES methods.
We further find that this is not only the case for the energy.
As we show in Fig~\ref{fig:obs_error_power} (a) and (b), we find that \black{measures} such as the quantum state fidelity $F(\sigma, \rho) = \tr[\sqrt{\sqrt{\rho}\sigma\sqrt{\rho}}]$ and trace distance $\|\rho - \sigma\|_1 = \frac{1}{2}\tr[\sqrt{(\rho - \sigma)^\dagger (\rho - \sigma)}]$ indicate that simulation can be done more accurately by constructing the power subspace. 
\black{Note that the trace distance gives upper bound on the deviation of an observable $O$ as $|\tr[\rho O] - \tr[\sigma O]| \leq \|\rho - \sigma\|_1 \|O\|_{\infty}$ where $\displaystyle \|O\|_\infty := \lim_{p \to \infty} \tr[(\sqrt{O^\dag O})^p]^{1/p}$, which is due to the tracial matrix Hoelder's inequality}. This leads us to expect that important physical quantities such as $k$-point correlators, e.g. two-point correlators $\braket{Z_0 Z_r}$  as shown in Fig.~\ref{fig:obs_error_power}(c), are better estimated when the trace distance is smaller. 
}

\subsection{Errors of physical observables under fault subspace}
\begin{figure}[t]
\begin{center}
\resizebox{0.99\columnwidth}{!}{\includegraphics{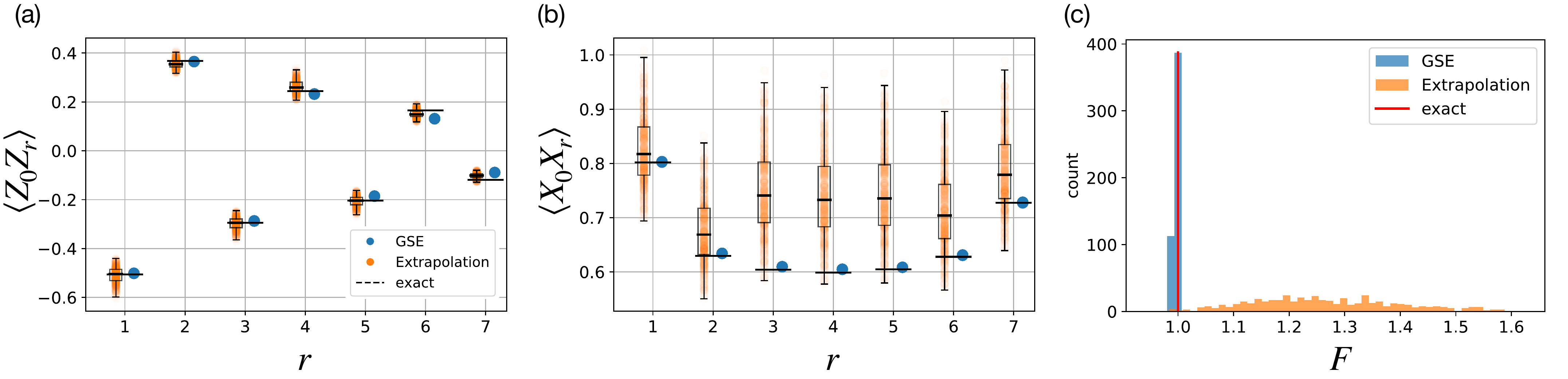}}
\caption{\black{Analysis on the performance of error-mitigated states that are effectively realized by the GSE method using the fault subspace and error-extrapolation method. (a)(b) Two-point correlators $\braket{Z_0 Z_r}$ and $\braket{X_0 X_r}$ where $r$ is the site index. 
(c) The quantum state fidelity $F$ between the error-mitigated states $\widetilde{\rho}$ and the exact ground state $\rho_{\rm GS}$. All data from the GSE method satisfies $F\lesssim 1$, which is violated in the error-extrapolation method.  We fix the error unit to be $\epsilon=1.5$, and generate 500 sets of noisy quantum states $\rho(\hat{\lambda}_i \epsilon)$ where $\hat{\lambda}_i = \lambda_i + \mathcal{N}(0, \lambda_i \epsilon\sigma^2)$ for $\lambda_i \in \{1, 2, 3\}$ and $\sigma = 0.1$. We assume that each Pauli term is estimated without any shot-noise.
The variational ansatz consists of depth $d=6$.}
}
\label{fig:obs_error_fault}
\end{center}
\end{figure}

\black{
We can also perform similar analysis for the fault subspace by comparing with the error-extrapolation method.
Figure~\ref{fig:obs_error_fault}(a) and (b) show the results for computing two-point correlators $\braket{Z_0 Z_r}$ and $\braket{X_0 X_r}$, respectively.
In both plots, we observe a huge separation in the variance, which reflects the vulnerability of the error-extrapolation method against the fluctuation in noise level control.
Even more prominent data is shown in Fig.~\ref{fig:obs_error_fault}(c).
We find that the effective quantum state $\rho_{\rm extrap} = \sum_i \beta_i \rho(\lambda_i \epsilon)$ is highly nonphysical, which results in fidelity $F(\rho_{\rm extrap}, \rho_{\rm GS}) > 1.$
In sharp contrast, we find that the effective state constructed by the fault subspace always satisfies $F \lesssim 1$. 
}

\section{Effect of shot noise on the generalized quantum subspace expansion} 
Here, we provide analytical results for the effect
of shot noise to the solution of the generalized eigenvalue problem~\cite{trefethen1997numerical}. 
First we introduce noise-free matrices that describe the quantum subspace;
let us denote the Hamiltonian in the expanded subspace and the overlap matrix denote as $\mathcal{H}_0$ and $\mathcal{S}_0$, respectively. 
Note that $\mathcal{S}_0 \geq 0$. The generalized eigenvalue problem without shot noise can be described as $\mathcal{H}_0 \vec{\alpha}_0=E_{00} \mathcal{S}_0 \vec{\alpha}_0$, where $\vec{\alpha}_0$ is the error-free ideal solution. In the following discussion, we assume the ideal eigenvalues are not degenerated. Denoting the effects of shot noise as $\delta \mathh$ and $\delta \maths$, we represent $\mathh=\mathh_0+\delta \mathh$ and $\maths=\maths_0+\delta \maths$.  Now, we have 
\begin{equation}
\mathh \vecc_n= E_n \maths \vecc_n.
\end{equation}
Here, $\vecc_n=\vecc_{on}+\delta \vecc_n$ and $E_n=E_{0 n}+\delta E_n$ with $E_{0n}$ and $\vecc_{0n}$ representing the ideal $n$-th eigenvalue and solution of the generalized eigenvalue problem, respectively. Focusing on the first order of the deviation, we get 
\begin{align}
\mathh_0  \delta \vecc_{n}+\delta \mathh \vecc_{0n}=E_{0n}\maths_0 \delta \vecc_n+E_{0n}\delta \maths \vecc_{0n}+\delta E_n \maths_0 \vecc_{0n}.
\label{Eq: deviation}
\end{align}
Now, by expanding 
$\delta \vecc_n=\sum_{m} \epsilon_{nm} \vec{\alpha}_{0m}$, we have
\begin{align}
\sum_m \epsilon_{nm} E_{0m} \maths_0 \vecc_{0m}+\delta \mathh \vecc_{0n}=E_{0n} \maths_0 \sum_m \epsilon_{nm} \vecc_{0m}+E_{0n}
 \delta \maths \vecc_{0n}+\delta E_n \maths_0 \vecc_{0n}
 \label{Eq: deviation2}
 \end{align}

By multiplying $\vecc_{0n}^{~\dag}$ from the left in Eq.~(\ref{Eq: deviation2}), we obtain 
\begin{align}
\delta E_n= \vecc_{0n}^{~\dag} (\delta \mathh - E_{0n}\delta \maths) \vecc_{0n},
\label{Eq: enrgydeviation}
\end{align}
where we used 
\begin{align}
\vecc_{0m}^{~\dag} \maths_0 \vecc_{0n}= \delta_{nm}. 
\label{Eq: orthonormal} 
\end{align}
Multiplying $\vecc_{0l}^{~\dag}~(l\neq n)$ from the left 
in Eq.~(\ref{Eq: deviation2})
leads to 
\begin{align}
\epsilon_{nl}= \frac{\vecc_{0l}^{~\dag} (\delta \mathh - E_{0n}\delta \maths)\vecc_{0n}}{E_{0n}-E_{0l}}.
\end{align}
We also note that from the normalization condition $\vecc_n ^\dag \maths \vecc_n=1$, we can derive $\epsilon_{nn}=-\frac{1}{2} \vecc_{0n}\delta \maths \vecc_{0n}$. To summarize, we obtain 
\begin{equation}
\begin{aligned}
E_n & = E_{0n}+\vecc_{0n}^{~\dag} (\delta \mathh - E_{0n}\delta \maths) \vecc_{0n} \\
\vecc_n &= (1-\frac{1}{2} \vecc_{0n}\delta \maths \vecc_{0n})\vecc_{0n}+\sum_{m\neq n} \frac{\vecc_{0m}^{~\dag} (\delta \mathh - E_{0n}\delta \maths)\vecc_{0n}}{E_{0n}-E_{0m}}\vecc_{0m}.
\end{aligned}
\end{equation}

Let us denote
the total number of measurements and the dimension of the subspace as $N_{\rm s}$  and $D$. 
Since each element of $\delta \mathh$ and $\delta \maths$ is in the order of $O(D N_{\rm s}^{-\frac{1}{2}})$, we have $\delta E_n=O(D N_{\rm s}^{-\frac{1}{2}})$ and $\delta \vecc_n=O(D N_{\rm s}^{-\frac{1}{2}})$. 
We can see $\{ S^{1/2}_0 \vec{\alpha}_n \}_n$ are mutually orthogonal and normalized due to Eq.~(\ref{Eq: orthonormal}). Thus, for an \black{arbitrary $D\times D$ matrix $B$}, we have
\begin{align}
|\vec{\alpha}_{0 n}^\dag B \alpha_{0 n}| = 
|\vec{\alpha}_{0 n}^\dag \maths_0^{1/2} \maths_0^{-1/2} B \maths_0^{-1/2} \maths_0^{1/2} \vec{\alpha}_{0 n}| \leq
 \| \maths_0^{-1/2} B \maths_0^{-1/2}  \|_{op},
\label{Eq:upperbound1}
\end{align}
where $\|\cdot \|_{op}$ denotes an operator norm. \black{Substituting $B=\delta \mathh - E_{0n} \delta \maths$ to Eq.~\eqref{Eq:upperbound1} results in}
\begin{equation}
\begin{aligned}
|\delta E_n| &\leq \|\maths_0^{-1/2} (\delta \mathh -E_{0n} \delta \maths) \maths_0^{-1/2} \|_{op} \leq \|\maths_0^{-1} \|_{op} (\|\delta \mathh \|_{op}+ |E_{0n}| \|\delta \maths \|_{op} ) \\
& \lesssim  \frac{2 D \|\maths_0^{-1} \|_{op}}{\sqrt {N_{\rm s}}} (\|\mathh \|_{F}+ |E_{0n}| \|\maths \|_{F}),
\end{aligned}
\end{equation}
where $\|\cdot\|_F$ is the Frobenius norm and we used $\|C\|_{op} \leq \|C\|_F$ for an arbitrary matrix $C$, and $\|\delta \mathh\|_F \lesssim \|\mathh\|_F/\sqrt{N_{\rm s} D^{-2}/2}$ and $\|\delta \maths\|_F \lesssim \|\maths\|_F/\sqrt{N_{\rm s} D^{-2}/2}$. 
Also, we assigned the same number of samples to the measurement of $\mathh$ and $\maths$. 
Here, we assumed $A=\rho$ or $A=I/d$ with $d$ being the system dimension. Note that rescaling the $A$ matrix does not affect the overall accuracy.

When we decompose the Hamiltonian into the linear combination of Pauli operators as $H=\sum_a h_a P_a$, we get $|\mathh_{ij}| \leq \gamma$ with $\gamma=\sum_a |h_a|$. 
Thus, we obtain
\begin{align}
|\delta E_n| \leq \frac{4 \gamma D^2 \|\maths_0^{-1} \|_{op}}{\sqrt{N_{\rm s}}}, 
\end{align}
where we used $E_{0n} \leq \|H\|_{op} \leq \gamma$, and also assumed $\maths_{ij} \leq 1$, which is satisfied in highly practical cases such as the power subspace (without additional bases) and fault subspace.
Thus, we can achieve the required accuracy $\varepsilon$ when $N_{\rm s} \geq 16 \gamma^2 D^4\|\maths_0^{-1} \|^2/\varepsilon^2$. 

\black{We further discuss $\|\maths_0^{-1}\|_{op}$ for the power subspace in the case of $D=2$ and $D=3$, which is highly practical for near-term devices.} For $D=2$,  when we set $A=\rho$ for the ansatz given in Eq.~\eqref{Eq:gse_ansatz}, we have
\begin{align}
\maths_0= \begin{pmatrix}
1 & \tr[\rho^2] \\
\tr[\rho^2] & \tr[\rho^3] \\
\end{pmatrix},
\end{align}
and we obtain $\|\maths_0^{-1}\|_{op} \approx (\tr[\rho^3]-(\tr[\rho^2])^2)^{-1}$ under the assumption of $(\tr[\rho^2])^2 \ll 1$. Similarly, we have $\|\maths_0^{-1}\|_{op} \approx (\tr[\rho^2])^{-1}$ for $A=I/d$. 
\textcolor{black}{In the case of $D=3$, we obtain $\|\maths_0^{-1}\|_{op} \approx (\tr[\rho^5]-(\tr[\rho^4])^2/\tr[\rho^3])^{-1}$ ($A=\rho$) and $\|\maths_0^{-1}\|_{op} \approx (\tr[\rho^4]-(\tr[\rho^3])^2/\tr[\rho^2])^{-1}$ ($A=I/d$), respectively.}

\section{Effect of shot noise on observable estimation}
\black{As is discussed in the previous section, the values of physical observable necessarily deviates from the ideal ones unless we take infinite number of measurements. Such fluctuations are often called the shot noise. }
For instance, results provided in Fig.~\ref{fig:gse} in the main text take the effect of shot noise into account.
Each matrix element $\mathcal{H}_{ij}$ (or $\mathcal{S}_{ij})$ is given as $\mathh_{ij} = \mathh_{ij}^{(0)} + \delta \mathh_{ij}$, where $\mathh_{ij}^{(0)}$ is the expectation value and $\delta \mathh_{ij}$ is a Gaussian noise whose amplitude is determined from the variance that arise from projective measurements.
In the following, we discuss the variance of estimators for physical observables such as $\mathh_{ij}$ and $\maths_{ij}$, which are computed by letting multiple quantum states interfere with each other.

\subsection{Power subspace}
It is instructive to start from a case where we have $M$ identical copies to compute the expectation value of an operator $O$. 
As was proposed in Ref.~\cite{huggins2020virtual, koczor2020exponential}, let us employ the cyclic shift operator $S^{(M)}$ which act on $M$ systems as follows,
\begin{equation}
    S^{(M)} \ket{\psi_1} \otimes \ket{\psi_2} \otimes \cdots \otimes \ket{\psi_M} = \ket{\psi_2} \otimes \ket{\psi_3} \otimes \cdots \otimes \ket{\psi_M} \otimes \ket{\psi_1},
\end{equation}
where $\ket{\psi_m}$ denotes an arbitrary quantum state of $m$-th system.
Practically, $S^{(M)}$ can be regarded as a product of shift operators between two systems as
$S^{(M)}= \prod_m S^{(2)}_{m, m+1}$
that can be constructed as a tensor product of swap operators acting on all corresponding qubits
where $S^{(2)}_{m, m+1} \ket{\psi_m} \otimes \ket{\psi_{m+1}} = \ket{\psi_{m+1}} \otimes \ket{\psi_m}$. 
After some computation, we can show that the expectation value for the power of quantum state $\rho^M$ can be expressed as
\begin{equation}
    \braket{O} = \frac{\tr[\rho^M O]}{\tr[\rho^M]}
    = \frac{\tr[S^{(M)} O^{(M)} \bigotimes_{m=1}^M \rho]}{\tr[S^{(M)} \bigotimes_{m=1}^M \rho]},\label{Eq:VD_expectation}
\end{equation}
where we take $O^{(M)} = O\otimes I \otimes \cdots \cdots I$ to provide a unified view over various choice of $M$~\cite{huggins2020virtual}. 
The variance for the single-shot estimation of the numerator can be given as 
\begin{eqnarray}
    {\rm Var}(S^{(M)} O^{(M)}) &=& \braket{\braket{(\Sm \Om)^2}} - \braket{\braket{\Sm \Om}}^2 \nonumber\\
    &=& \braket{\braket{\Sm \Om \Sm \Om}} - \tr[\rho^M O]^2 \nonumber\\
    &=& \braket{\braket{(I\otimes O \otimes I \otimes \cdots \otimes I ) (O \otimes I \otimes \cdots \otimes I)}} - \tr[\rho^M O]^2 \nonumber\\
    &=& \tr[\rho O]^2 - \tr[\rho^M O]^2,\label{Eq:var_numerator}
\end{eqnarray}
where $\braket{\braket{\cdot}} := \tr[(\cdot) \otimes_{m=1}^M \rho]$ is introduced to denote the expectation value computed in the extended Hilbert space, which consists of $M$ systems. 
It is beneficial to remark on a case when the observable is a linear combination of Pauli operators; we decompose as 
$O = \sum_{k=1}^{N_{\rm{O}}} c_k P_k$ where $P_k$ is a product of Pauli operators and $N_{\rm{O}}$ denotes the number of terms. Then, the variance is given as
\begin{eqnarray}
    {\rm Var}(\Sm \Om) = \sum_k |c_k|^2 {\rm Var}(\Sm (P_k^{(M)})).\label{eq:SO_var_sum}
\end{eqnarray}
Here, we assume that we measure ${\rm{Tr}}[\rho^M P_1]$,
${\rm{Tr}}[\rho^M P_2]$, $\cdots$, ${\rm{Tr}}[\rho^M P_{N_{\rm{O}}}]$  in separate experiments, and therefore the total variance~\ref{eq:SO_var_sum} is estimated as a sum of each term.
The variance for the single-shot estimation of the denominator can be obtained from a parallel argument as 
\begin{eqnarray}
    {\rm Var}(\Sm) &=& \braket{\braket{(\Sm)^2}} - \braket{\braket{\Sm}} ^2 \nonumber\\
    &=& 1 - \tr[\rho^M]^2.\label{Eq:var_denominator}
\end{eqnarray}
Finally, we can calculate the variance of the estimator of the observable itself  by substituting the above expressions into the standard formula. 
If we take $n_{\rm s}$ measurement shots for every term, the overall variance can be calculated as
\begin{eqnarray}
    \frac{{\rm Var}(O)}{n_{\rm s}} 
    {\rm{\simeq}}
    \frac{\braket{O}^2}{n_{\rm s}}
    \left(\frac{{\rm Var}(\Sm \Om)}{\bbraket{\Sm \Om}^2} + \frac{{\rm Var}(\Sm)}{\bbraket{\Sm}^2} 
    - \black{2\frac{{\rm Covar}(\Sm \Om, \Sm)}{\bbraket{S^{(M)}O^{(M)}} \bbraket{S^{(M)}}}
    } \right),\label{Eq:O_variance}
\end{eqnarray}
where the covariance between $\Sm\Om$ and $\Sm$ is set to zero in our numerical simulation. 
This is because we have chosen $\Om$ to be $[S^{(M)}O^{(M)}, S^{(M)}] \neq 0$. In this case, the denominator and numerator of Eq.~\eqref{Eq:VD_expectation} cannot be measured simultaneously, and hence we can ignore the covariance by measuring them separately.
\black{Let us also remark that overall variance can be improved, e.g., by weighing the number of measurements according to the coefficient $c_k$.}

We have used Eq.~\eqref{Eq:O_variance} to estimate the variance of the matrix elements $\mathcal{H}_{ij}$ and $\mathcal{S}_{ij}$ for the power subspace.
Note that \black{bases of subspaces $\{\sigma_i\}$, including the additional terms $\rho^m H$}, are chosen so that all elements are given as $\tr[\rho^m H^k]~(k\in\mathbb{Z}^{+})$.
We later discuss the case when we must compute, e.g., $\tr[\rho H \rho H].$

\subsection{Fault subspace}
Next, we discuss the case when the error-mitigated quantum state is expressed by non-identical quantum states. 
It is practical to first set $M=2$ so that we can describe the case for the fault subspace.
Given a quantum state $\rho_{\rm EM} = \sum_{i,j=0}^m \alpha_i^* \alpha_j \rho_i \rho_j$
with $\rho_i=\rho(\lambda_i \epsilon)$ and $M=2m$, the expectation value of a physical observable $O$ is expressed as
\begin{eqnarray}
    \braket{O} = \frac{\tr[\rho_{\rm EM} O]}{\tr[\rho_{\rm EM}]} = 
    \frac{\sum_{ij}\alpha_i^* \alpha_j \tr[\rho_i \rho_j O]}{\sum_{ij} \alpha_i^* \alpha_j \tr[\rho_i \rho_j]} = 
    \frac{\sum_{ij}\alpha_i^* \alpha_j\tr[S^{(2)} O^{(2)} (\rho_i\otimes \rho_j)]}{\sum_{ij} \alpha_i^* \alpha_j \tr[S^{(2)}(\rho_i\otimes\rho_j)]} = \frac{\sum_{ij} \alpha_i^* \mathcal{O}_{ij} \alpha_j}{\sum_{ij} \alpha_i^* \mathcal{S}_{ij} \alpha_j},
\end{eqnarray}
where we have defined $\mathcal{O}_{ij} = \tr[\rho_i \rho_j O]$.
Based on the discussion of Eq.~\eqref{Eq:var_numerator}, we obtain the variance for the single-shot estimation of an element $\mathcal{O}_{ij}$ as
\begin{eqnarray}
    {\rm Var}(S^{(2)} O^{(2)}
    ) _{ij} &\equiv& \bbraket{(S^{(2)} O^{(2)})^2}_{ij} - \bbraket{S^{(2)} O^{(2)}}^2_{ij} \nonumber\\
    &=& \bbraket{(I\otimes O)(O\otimes I)}_{ij} - \tr[\rho_i \rho_j O]^2\nonumber \\
    &=& \tr[\rho_i O]\tr[\rho_j O] - \tr[\rho_i \rho_j O]^2,\label{Eq:var_Oij}
\end{eqnarray}
where we have defined $\bbraket{\cdot}_{ij} = \tr[(\cdot) (\rho_i\otimes \rho_j)]$. Also, in a similar way to Eq.~\eqref{Eq:var_denominator}, the variance for single-shot estimation of $\mathcal{S}_{ij}$ is given as
\begin{eqnarray}
   {\rm Var}(S^{(2)} 
    ) _{ij} &\equiv& \bbraket{(S^{(2)})^2}_{ij} - \bbraket{S^{(2)}}^2_{ij} \nonumber\\
    &=& 1 - \tr[\rho_i \rho_j]^2.\label{Eq:var_Sij}
\end{eqnarray}

Using the above expressions, we find that the variance under $n_{\rm s}$  measurement shots for each term can be formally given as
\begin{eqnarray}
    \frac{{\rm Var}(O)}{n_{\rm s}} = \frac{\braket{O}^2}{n_{\rm s}}\left(
    \frac{\sum_{ij} 
     (\alpha^*_i)^2 (\alpha_j)^2
    {\rm Var}(S^{(2)} O^{(2)}
    ) _{ij}
    }{\tr[\rho_{\rm EM}O]^2} + \frac{\sum_{ij} (\alpha^*_i)^2 (\alpha_j)^2
      {\rm Var}(S^{(2)} 
    ) _{ij}
    }{\tr[\rho_{\rm EM}]^2}
    \right).\label{Eq:O_variance_fault}
\end{eqnarray}

In the main text, we calculate the variance of the physical observable (in particular the energy $\braket{H}$) in a stochastic way rather than calculating from Eq.~\eqref{Eq:O_variance_fault}. 
This procedure can be summarized as follows: (1) add a Gaussian noise to $\mathcal{H}_{ij}$ and $\mathcal{S}_{ij}$ whose variance is determined from Eqs.~\eqref{Eq:var_Oij} and~\eqref{Eq:var_Sij}, (2) regularize $\mathcal{S}$ to omit small/negative eigenvalues, (3) solve the generalized linear equation, and (4) repeat (1)-(3) until the variance can be estimated with sufficient accuracy.

\subsection{General case}
Here, we briefly comment on the variance using the most general subspace.
For simplicity, we focus on the variance of estimating $\braket{O}_Q:=\tr[O(Q_1 \rho_1 Q_2 \rho_2 \cdots Q_M \rho_M)]$, where $Q_m~(m=1,...,M)$ is taken as a local Pauli operator.
This can be measured using the cyclic shift operator $\Sm$ and $\Om := O \otimes I \otimes \cdots \otimes I$
\begin{eqnarray}
    \braket{O}_Q = \tr[\Sm \Om (Q_1\otimes \cdots \otimes Q_M)(\rho_1\otimes \cdots \otimes \rho_M)],
\end{eqnarray}
and hence the variance of the estimator can be given as follows,
\begin{eqnarray}
    {\rm Var}(O)_Q &=& \bbraket{\left(\Sm \Om(\bigotimes_{m=1}^M Q_m)\right)^2} - \left(\bbraket{\Sm\Om \bigotimes_{m=1}^M Q_m}\right) ^2 \nonumber \\
    &=& \bbraket{\left(O Q_1\otimes Q_2 \otimes \cdots \otimes Q_M \right)
    \left(Q_2 \otimes Q_3 \otimes \cdots \otimes Q_M \otimes O Q_1 \right)}
    - \braket{O}_Q^2 \nonumber \\
    &=& \tr[\rho_1 O Q_1 Q_2]\tr[\rho_2 Q_2 Q_3] \cdots \tr[\rho_M Q_M O Q_1] - \bbraket{O}_Q^2,
\end{eqnarray}
where here we have denoted $\bbraket{\cdot} := \tr[(\cdot) \bigotimes_{m=1}^M \rho_m]$.

\section{Details regarding the structure of variational quantum circuits}\label{sec:hardware_efficient}
\begin{figure}[h]
\begin{center}
\resizebox{0.35\hsize}{!}{\includegraphics{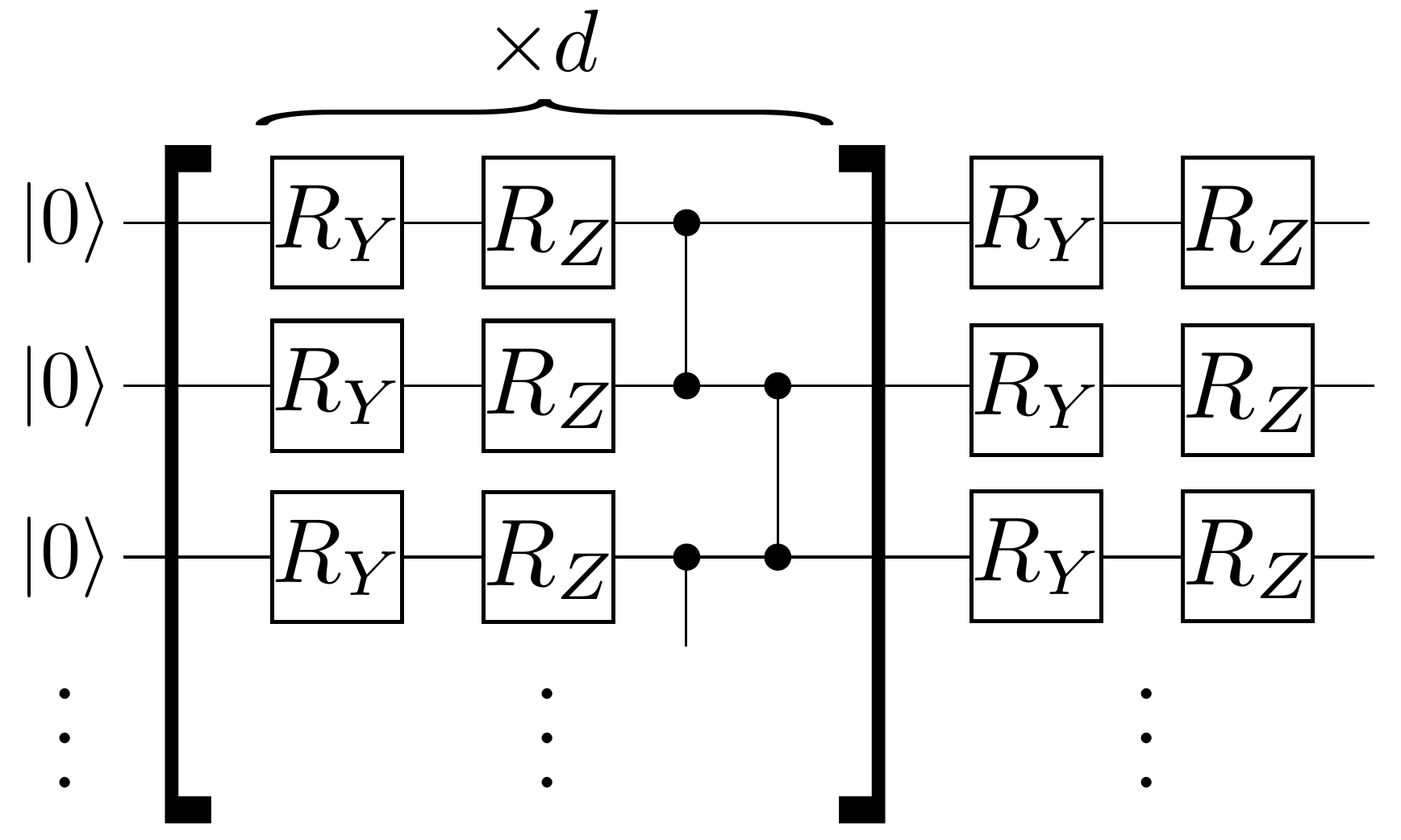}}
\caption{Structure of hardware-efficient ansatz employed in this work.
}
\label{fig:circuit}
\end{center}
\end{figure}

In the main text, we have simulated the ground state of the one-dimensional transverse-field Ising model $H = - \sum_r Z_r Z_{r+1} + h\sum_r X_r$ where $X_r$ and $Z_r$ denote the $x$- and $z$-component of the Pauli matrices acting on the $r$-th site under open boundary condition.
Throughout this paper, we have employed the hardware-efficient ansatz structure as shown in Fig.~\ref{fig:circuit}, where the repetition number of units has been taken as $d=12$ for results in Figs.~\ref{fig:gse} and ~\ref{fig:fse}, and $d=6$ for that in Fig.~\ref{fig:gse_pdep}.
Noisy quantum states are generated by adding local depolarizing noise after each quantum gate.

\section{Error-extrapolation method}
Here, we briefly review the error-extrapolation method used for the quantum error mitigation~\cite{temme2017error}.
Let us assume that a quantum state can be characterized by an error-control parameters as $\rho(\lambda \epsilon)$, where $\epsilon$ is a fixed value which can be considered as the minimum achievable noise level and $\lambda \geq 1$ is the stretch factor.
The expectation value of any observable $O$ will be expressed  as a power series around its noise-free value $O^*$ as 
\begin{equation}
    \langle O(\lambda \epsilon) \rangle = O^* + \sum_{k=1}^n a_k \lambda^k + \mathcal{O}(\lambda^{n+1}). \label{Eq:error_expansion}
\end{equation}
The noise-free value $O^*$ can be calculated from $n+1$ results corresponding to stretch factors $\{\lambda_i\}_{i=0}^n$ as
\begin{eqnarray}
O^* = \sum_{i=0}^n \beta_i \braket{O(\lambda_i \epsilon)} + \mathcal{O}(\epsilon^{n+1}),\label{Eq:extrapolation}
\end{eqnarray}
where $\beta_i = \prod_{j \neq i}\lambda_j(\lambda_j -\lambda_i)^{-1}$ are solutions of linear equation 
\begin{eqnarray}
\sum_i \beta_i = 1, \sum_i\beta_i \lambda_i^k = 0.
\end{eqnarray}
Since $\{\beta_i\}$ is common among any physical observable, this is equivalent to estimate the zero-noise density matrix as $\rho_{\rm ex} = \sum_i \beta_i \rho(\lambda_i \epsilon)$.
We remark that Eq.~\eqref{Eq:extrapolation} can be used  as long as $\braket{O(\lambda \epsilon)}$ is controlled by $\lambda$. 
Namely,  the error-extrapolation method can be applied not only for the calculation of the raw expectation value but also for the calculation with
the VD/EES method such as $\tr[\rho^M(\lambda_i \epsilon) H]/\tr[\rho^M(\lambda_i \epsilon)]$.

\begin{figure}[t]
\begin{center}
\resizebox{0.4\hsize}{!}{\includegraphics{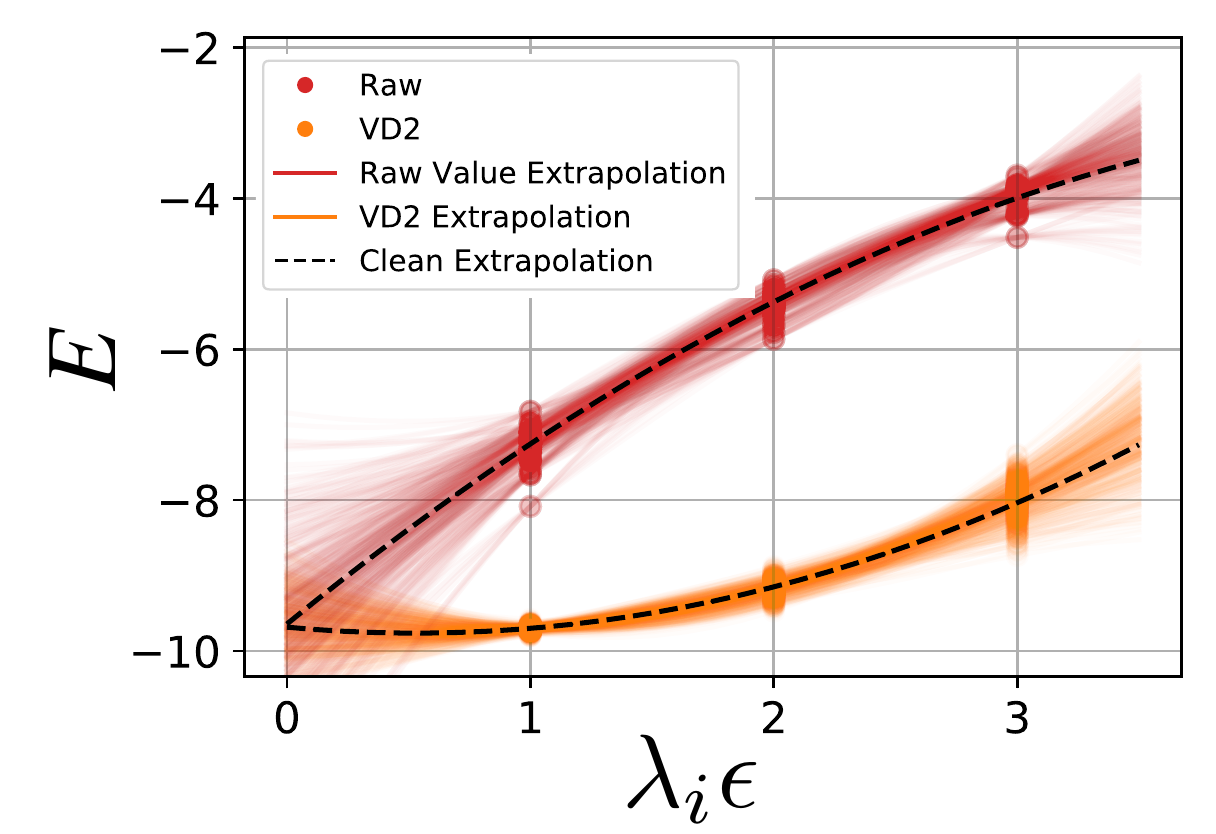}}
\caption{Error-extrapolation method to estimate the zero-noise limit of the ground-state energy. Red and orange dots denote the estimated energy from raw quantum states and VD/EES method using $M=2$ copies, respectively. The colored lines are the extrapolation fit for noisy values to obtain the noise-free values, and the black dotted line indicates extrapolation under perfect noise control.
For each error unit $\epsilon$, we generate 500 sets of noisy quantum states $\rho(\hat{\lambda}_i \epsilon)$ where $\hat{\lambda}_i = \lambda_i + \mathcal{N}(0, \lambda_i \sigma^2)$ for $\lambda_i \in \{1, 2, 3\}$ and $\sigma = 0.1$. 
}
\label{fig:extrapolation}
\end{center}
\end{figure}

\begin{figure}[t]
\begin{center}
\resizebox{0.95\hsize}{!}{\includegraphics{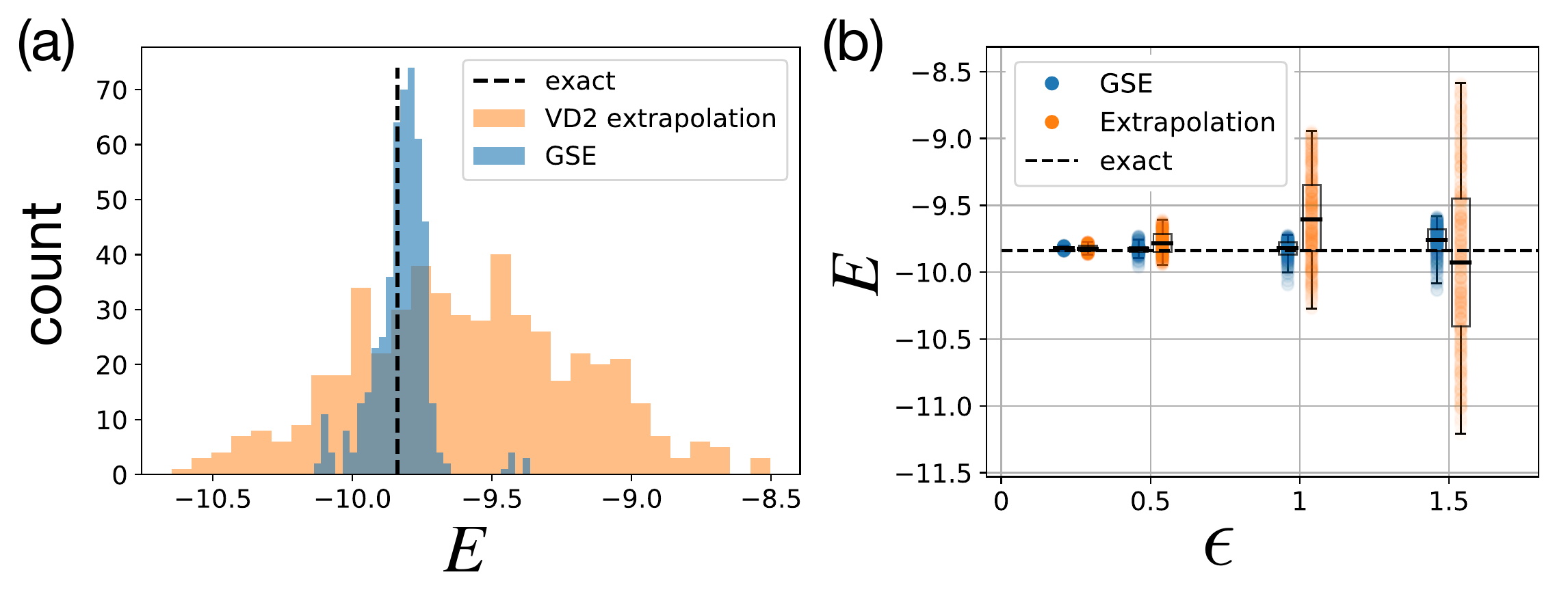}}
\caption{(a) Zero-noise energy estimation using the GSE method via fault subspace and the extrapolation of VD/EES method using $M=2$ copies. The unit of controlled noise parameter is taken as $\epsilon = 1$.
\black{(b) Comparing the effect of fluctuation in the stretch factor $\lambda$ under various $\epsilon$. The only difference from the calculation done for Fig.~\ref{fig:fse} in the main text is the effect of the shot-noise.
In all calculations, we consider the one-dimensional transverse-field Hamiltonian with  $N=8$ qubits with $h=1$, and we estimate each Pauli terms using $10^7$ measurement shots. }
}
\label{fig:fault_subspace_shotnoise}
\end{center}
\end{figure}

While the error-extrapolation technique is highly practical, it is crucial to perform perfect control over the stretch factor $\lambda$. 
This is demanding for the actual experimental setup, and therefore it is highly possible that $\lambda$ deviates from the actual target value.
Figure~\ref{fig:extrapolation} shows that the effect of such an unwanted fluctuation can be somewhat suppressed by employing data points calculated from the VD/EES method. 
However, it is evident from Fig.~\ref{fig:fault_subspace_shotnoise}(a) that the GSE method can estimate the error-free observable much more efficiently because such an effect barely affects the construction of the quantum subspace.
\black{This can be confirmed to be true for various error levels, especially when we take a sufficient number of measurement shots $n_s$ so that the effect of shot-noise is negligible (See Fig.~\ref{fig:fse} in the main text for a case with $n_s\rightarrow \infty$). 
When $n_s$ is finite, the result by the GSE method may not necessarily satisfy $E_{\rm GSE} \geq E_{\rm GS}$ where $E_{\rm GS}$ is the exact ground-state energy (See Fig.~\ref{fig:fault_subspace_shotnoise}(b)). However, we numerically confirm that the fluctuation is much smaller and hence is more reliable.
}

\black{It must be noted that the GSE method using the fault subspace can outperform the conventional error-extrapolation method even when the stretch factor $\lambda_i$ is precisely known. 
Let us consider distilling the error-extrapolated density matrix $\rho_{\rm ex}$ using $M=2$ copies as $\rho_{\rm ex}^2/\tr[\rho_{\rm ex}^2]$.}
\black{We can see that this state is included in the variational ansatz employed in the GSE method , i.e.,  $\rho_{\rm EM} = \sum_{ij} \alpha_i^* \alpha_j \rho(\lambda_i \epsilon) \rho(\lambda_j \epsilon)$.
When $\rho_{\rm ex}$ is a physical state, this also holds for higher orders since we can always construct a variational ansatz that includes $\rho_{\rm ex}^M/\tr[\rho_{\rm ex}^M]$. For instance, we can simply take $\sigma_i = \rho(\lambda_i \epsilon)$ and $A = \rho_{\rm ex}^{M-2}$ which yields $\rho_{\rm EM} = \sum_{ij} \alpha_i^* \alpha_j \rho(\lambda_i \epsilon) \rho_{\rm ex}^{M-2} \rho(\lambda_j \epsilon)$.
}

\end{document}